\documentclass{jfm}
\usepackage{graphicx}
\usepackage{amsmath}
\usepackage{amssymb}
\usepackage{xcolor}
\usepackage{epstopdf, epsfig}
\usepackage{multirow}
\usepackage{booktabs}
\usepackage{hyperref}
\usepackage{longtable}

\shorttitle{On the intracyclic instability in Stokes layers}
\shortauthor{M. Zhang}

\title{On the intracyclic instability in Stokes layers}

\author{
  Mengqi Zhang\aff{1}
  \corresp{\email{mpezmq@nus.edu.sg}}
}

\affiliation{
\aff{1}Department of Mechanical Engineering, National University of Singapore, 9 Engineering Drive 1, 117575, Republic of Singapore
}

\begin{document}

\maketitle

\begin{abstract}
Time-dependent fluid dynamics plays a crucial role in both natural phenomena and industrial applications. Understanding the flow instabilities and transitions within these dynamical systems is essential for predicting and controlling their unsteady behaviour. A classic example of time-dependent flow is the Stokes layer. To study the transition mechanism in this flow, we employ the Finite-Time Lyapunov Exponent (FTLE) to demonstrate that a linear energy amplification mechanism may explain the intracyclic instability in the transitional Stokes layer, supported by favourable comparisons with experimental measurements of axial turbulence intensity. This complements existing theories applied to the Stokes layer in the literature, including the Floquet analysis and the instantaneous/momentary analyses, which have struggled to capture this experimental observation accurately.  The FTLE analysis is closely related to the transient growth analysis, formulated as an optimisation problem of the disturbance energy growth over time. We found that the energy amplification weakens as the finite Stokes layer becomes more confined and the oscillating frequency has a non-monotonic effect on the maximum transient growth. Based on these results, we recommend future experimental studies to validate this linear mechanism.
\end{abstract}


\section{Introduction}
Time-dependent systems, including periodic, quasi-periodic, and chaotic dynamical systems, can exhibit complex behaviours. These behaviours are often more intricate compared to those of autonomous systems. By studying time-dependent flows, engineers and scientists can better predict and control the fluid dynamics in oscillatory flow systems, leading to improved designs and more efficient processes across multiple disciplines, including rheology characterisation, wave energy conversion, modelling of pulsatile blood flows and medical diagnostics. One of the fundamental time-dependent systems in fluid mechanics is the Stokes layer, a thin fluid layer near a solid boundary subject to a periodic motion, named after Sir George Gabriel Stokes. Analytical solutions exist for this flow and stability analyses of these solutions have been conducted extensively in the past due to their theoretical significance. However, the physical mechanism underpinning the flow transition in Stokes layers still remains poorly understood \citep{Davis1976}. This study aims to address the longstanding discrepancy between theoretical predictions and experimental observations in transitional Stokes layers by investigating their intracyclic instability.

As a time-periodic flow system, the linear dynamics of the Stokes layer has been initially studied using the Floquet theory by  \cite{Kerczek1974} and \cite{Hall1978}. Leveraging more powerful computational resources, \cite{Blennerhassett2002,Blennerhassett2006} were the first to identify an unstable Floquet mode in semi-infinite Stokes layers, determining the critical Reynolds number to be $Re_c\approx 708$, which is defined based on the Stokes layer thickness. However, experimental studies on the transitional Stokes layers report a broad range of transition Reynolds numbers, from 140 to 300 \citep{Hino1976,Hino1983,Jensen1989,Akhavan1991}, significantly lower than the theoretical prediction. This signals a typical subcritical transition scenario. 

While Floquet theory has provided insights into the long-term behaviour of disturbances in time-periodic systems, this approach falls short in capturing the intracyclic instability of the Stokes layer observed experimentally with significant modulation in each oscillation cycle. This has led to the study of instability using instantaneous or momentary stability theory \citep{Cowley1987a,Luo2010,Blondeaux2021}, which assesses the local stability by treating the base flow profile as frozen at each moment in time and assuming that the disturbance is of high frequency. The theory typically predicts disturbance decay at the start of the deceleration phase of the wall motion, which conflicts with experimental observations. On the other hand, using the non-normal stability theory, \cite{Biau2016} found large transient energy growth due to the two-dimensional (2-D) Orr mechanism in subcritical semi-infinite Stokes layers. In experiments, \cite{Akhavan1991a} have already observed the transient growth in their experiments, even though they connected their observations to the quasi-steady theory. The non-normal property of the Stokes layers is also evidenced in the exceptionally large flow response investigated by \cite{Blennerhassett2002} and \cite{Thomas2010} and the influence of high-frequency noise on the flow explored by \cite{Thomas2015}. 

Additional factors have been taken into account to further examine the transitional process in the Stokes layer. For example, (weakly) nonlinear stability theories were developed by \cite{Wu1992,Monkewitz1985} to study the nonlinear interaction of the salient modes in the flow. Besides, the role of wall roughness in the transition has been elucidated by \cite{Blondeaux1994,Vittori1998,Luo2010}. However, compelling evidence for a meaningful comparison with experimental results is still lacking. As a result, it remains unclear which theory is most relevant to the flow dynamics in Stokes layers from a practical standpoint. A significant research gap remains in our understanding of how the Stokes layer becomes unstable and transitions to turbulence.

To further clarify the flow instability and transition mechanism in the Stokes layer, this study revisits the linear dynamics of the periodic flow, aiming to determine whether the theory aligns with experimental results. Our focus is on the transient intracyclic instability, analyzed through the Finite-Time Lyapunov Exponent (FTLE) framework. Unlike the instantaneous or momentary stability theory, the FTLE analysis inherently incorporates the flow’s evolutionary history, without requiring a frozen base-flow profile. It also connects the transient growth and the numerical abscissa at short times and the Floquet exponent at long times, reconciling and complementing the existing theories. The obtained results in general align with prior experimental and numerical observations.

In the following, we will first formulate the linear equations and then introduce the numerical method to study these equations in Section \ref{Methodology}. Section \ref{Results} will show the results. We first illustrate the finite Stokes layer with $h=5$ (where $h$ is the nondimensionalised channel half-height, to be defined), corresponding to the case with the maximum transient growth. Then the $h=10$ case will be analysed and compared to the experimental data. Stokes layers with small-$h$ will be also studied. This is followed by an investigation of the effect of the oscillation frequency, the most practical parameter to vary in experiments, on transient growth. Additionally, nonlinear simulations will be presented to examine both the transient evolution and the saturated dynamics of the Stokes layer. Finally, we conclude the paper with some discussions that underscore the favourable comparison to experimental observations, establishing the non-normal mechanism as the key ingredient for the intracyclic instability in the Stokes layer, and outlining future directions for flow control and nonlinear analysis.

\section{Problem formulation and numerical methods}\label{Methodology}
\subsection{Formulation}\label{Equation}

We consider a finite Stokes layer in a channel, subject to a harmonic motion of the two oscillating walls in their own planes. The walls move at the same velocity $U_0\cos\omega \hat t$, where $U_0$ is the maximum oscillating velocity, and $\omega$ denotes the oscillating frequency. The mark hat indicates dimensional variables. The two walls are separated by $2d$ with the Cartesian coordinates located at the channel centre. For sufficiently distanced walls, this flow mimics the dynamics of the semi-infinite Stokes layer. The streamwise, wall-normal and transverse directions are denoted by $(\hat x,\hat y,\hat z)$, respectively, corresponding to the velocity components $\hat u,\hat v,\hat w$ in the three directions. The streamwise and transverse wavenumbers are represented by $\alpha$ and $\gamma$, respectively.

Following \cite{Blennerhassett2002,Blennerhassett2006}, we non-dimensionalise the flow system using the Stokes layer thickness $\sqrt{2\nu/\omega}$, the maximum wall-oscillating velocity $U_0$, the time scale $1/\omega$, and the pressure scale $\rho U_0^2$, where $\nu$ is the kinematic viscosity coefficient and $\rho$ is the density. The non-dimensional incompressible Navier-Stokes (NS) equations read
\begin{align}\label{eq21}
\frac{\partial \tilde{\mathbf u }}{\partial t} + Re (\tilde{\mathbf u}\cdot \nabla )\tilde{\mathbf u} = -Re\nabla \tilde p + \frac{1}{2}\nabla^2 \tilde{\mathbf u}, \ \ \ \nabla\cdot\tilde{\mathbf u}=0,
\end{align}
where the Reynolds number is defined as 
\begin{align}
Re=\frac{U_0}{\sqrt{2\nu\omega}} =\frac{1/\omega}{\sqrt{2\nu/\omega}/U_0}.
\end{align}
The second equation implies that the $Re$ can be thought of as the ratio between the time scale for wall oscillation and the time scale for the `penetration' effect of the wall oscillation. No-slip boundary conditions for the velocity components are imposed on the channel walls. Driven by the nondimensionalised periodic wall motion $\cos t$, the equations admit an analytical solution of a time-periodic flow $\mathbf U_b$ in the $x$ direction, i.e., 
\begin{align}
{\mathbf U_b}=(U_b(y,t),0,0) = (U_1(y) e^{i t} + U_1^*(y)e^{-i t} , 0 ,0 )   \ \ \ \ \ \text{with} \ \ \ \ \ U_1(y) = \frac{   \cosh{\sqrt{2i}y}   }{  2\cosh\sqrt{2i}h }  
\end{align}
where $h = d/\sqrt{2\nu/\omega}$ denotes the nondimensional channel half-height, $i$ is the imaginary unit and the superscript $^*$ marks the complex conjugate.

The linear analysis is formulated based on the decomposition of the flow variables into a sum of a base flow plus a fluctuation component, i.e., $\tilde{\mathbf u}({\mathbf x},t)={\mathbf U_b}({\mathbf x}, t) + {\mathbf u}({\mathbf x},t), \tilde p({\mathbf x},t)=p({\mathbf x},t)$, where there is no base pressure gradient. By substituting these equations into Eq. (\ref{eq21}), expanding all the terms, subtracting the base-flow terms and neglecting the nonlinear terms, we obtain
\begin{subeqnarray}\label{lineareq}
\frac{\partial  {\mathbf u}}{\partial t} + Re[ {\mathbf U_b} \cdot \nabla  {\mathbf u} +  {\mathbf u} \cdot \nabla {\mathbf U_b}  ] &=& -Re\nabla p + \frac{1}{2}\nabla^2  {\mathbf u},  \\ 
\nabla \cdot {\mathbf u}&=&0.
\end{subeqnarray}
The initial value problem Eq. (\ref{lineareq}) can be recast as $\frac{\partial  \mathbf u}{\partial t}=\mathbf L(t)\mathbf u$. In our calculation, the pressure term in the momentum equation was eliminated using the continuity condition, leading to the $v-\eta$ formulation based on the vertical velocity $v$ and the vertical vorticity $\eta= \frac{\partial u}{\partial z} - \frac{\partial w}{\partial x}$. The $v-\eta$ formulation reads
\begin{align}\label{eq13}
\frac{\partial  }{\partial t} \begin{bmatrix}
v \\
\eta
\end{bmatrix} &=  \begin{bmatrix}
\nabla^2 & 0 \\
0 & 1
\end{bmatrix} ^{-1} \begin{bmatrix}
 L_{11}(t) & 0 \\
 L_{21}(t)   &  L_{22}(t)
\end{bmatrix} \begin{bmatrix}
v \\
\eta
\end{bmatrix}  \\ 
 L_{11}(t)&=-ReU_b(t) \frac{\partial }{\partial x}   \nabla^2+ ReU_b''(t)\frac{\partial }{\partial x} +   \frac{1}{2}\nabla^4  \nonumber\\
 L_{21}(t)&=- ReU'_b(t)\frac{\partial }{\partial z}, \ \ \ 
 L_{22}(t)=-ReU_b(t) \frac{\partial  }{\partial x} +  \frac{1}{2}\nabla^2  \nonumber
\end{align}
where prime $'$ denotes the $y-$derivative. The boundary conditions of the perturbed variables are $v(\pm h)=v'(\pm h)=\eta(\pm h)=0$. 

To numerically discretise these equations, the classic spectral collocation method \citep{Weideman2000} is used for the spatial discretisation with a grid resolution $N_y=69$ (excluding the boundary grid points) in the wall-normal direction. The homogeneous boundary conditions of the variables are implemented by removing the first/last rows and first/last columns of the corresponding matrices in the collocation method. The clamped boundary conditions of $v$ are implemented using the code scripts provided by \cite{Weideman2000}. The fourth-order backward Euler method is used for time integration with $dt=10^{-5}$ in solving the linear equations.

\subsection{The FTLE analysis}\label{Method}

The FTLE analysis \citep{Haller2015,Shadden2012,Lekien2007} is traditionally applied to nonlinear flow systems to probe how initially close trajectories are separated with time. The maximum FTLE $\Lambda$ is defined as  $\boldsymbol{u}$, $\mathbf{u}$, $\boldsymbol{u}$
\begin{align}
\| \boldsymbol u(t)\|_2 \approx e^{2\Lambda (t-t_0)} \|\boldsymbol u(t_0)\|_2   \ \ \ \ \text{or}\ \ \ \ \frac{\| \boldsymbol u(t)\|_2}{\| \boldsymbol u(t_0)\|_2}\approx e^{2\Lambda (t-t_0)}.
\end{align}
In general, using the flow map $F^t_{t_0}$ to represent a (nonlinear) trajectory $\boldsymbol {\tilde u}$ from $t_0$ to $t$, the dynamics of the perturbation to $\boldsymbol {\tilde u}(t)$, denoted as $\boldsymbol u(t)$, can be approximated as
 \begin{align}
 \boldsymbol {u}(t)= F^t_{t_0}(\boldsymbol {\tilde u}(t_0)+\boldsymbol { u}(t_0)) - F^t_{t_0}(\boldsymbol {\tilde u}(t_0)). 
 \end{align}
 By Taylor expansion, $ \boldsymbol {u}(t)=\nabla F^t_{t_0}(\boldsymbol {\tilde u}(t_0)) \boldsymbol { u}(t_0)$, where $\nabla F^t_{t_0}(\boldsymbol {\tilde u}(t_0))$ is called the deformation gradient. The energy norm of $\boldsymbol {u}(t)$ can then be expressed as 
 \begin{align}
\| \boldsymbol u(t)\|_2 = \boldsymbol { u}(t_0)^*\nabla {F^t_{t_0}}^*(\boldsymbol {\tilde u}(t_0))  \nabla F^t_{t_0}(\boldsymbol {\tilde u}(t_0)) \boldsymbol { u}(t_0).
  \end{align}
 The energy ratio $\| \boldsymbol u(t)\|_2/\| \boldsymbol u(t_0)\|_2$ represents the Rayleigh quotient of the right Cauchy–Green strain tensor $\nabla {F^t_{t_0}}^*(\boldsymbol {\tilde u}(t_0))  \nabla F^t_{t_0}(\boldsymbol {\tilde u}(t_0)) $, or the squared first singular value of $\nabla F^t_{t_0}(\boldsymbol {\tilde u}(t_0)) $, which is quantitatively equal to the induced 2-norm $\| \nabla F^t_{t_0}(\boldsymbol {\tilde u}(t_0))  \|_2$. The 2-norm, defined as $\|\mathbf u\|_2=\frac{1}{2}\int_V u^*u + v^*v + w^*w\ dV$, calculates the kinetic energy density in the flow. 

\begin{figure}
	\centering
	\includegraphics[width=0.48\textwidth,trim= 40 195 80 0,clip]{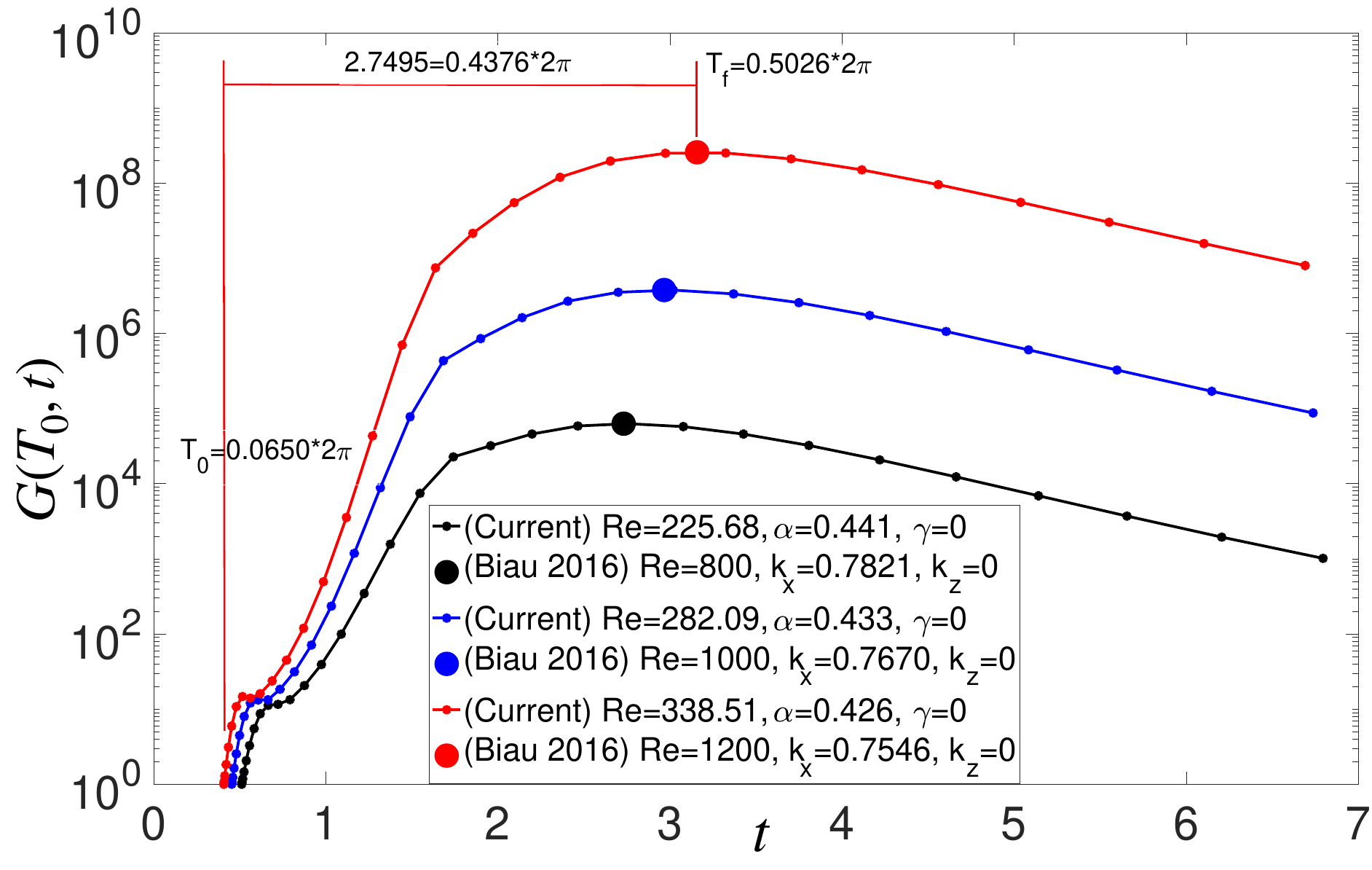}		
	\put(-194,110){$(a)$}	\ \ \ \ \ 
	\includegraphics[width=0.48\textwidth,trim= 10 125 20 15,clip]{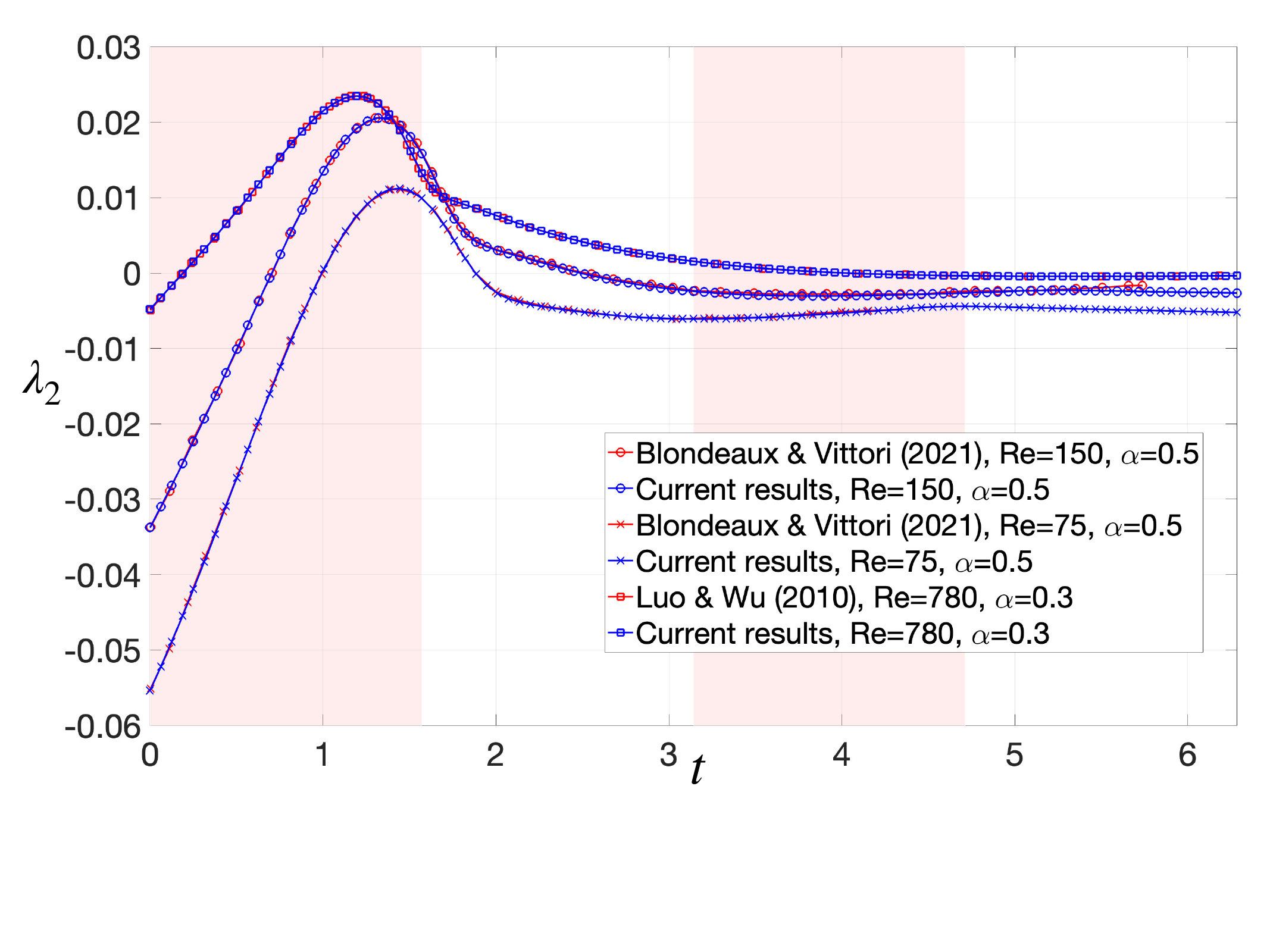}
	\put(-188,110){$(b)$}	\	
	\caption{($a$) Transient growth in 2-D Stokes layers with parameters identical to those in \cite{Biau2016}. The two studies use different non-dimensionalisation methods for the flow system, necessitating conversion of the parameters; see the legend for details. The finite domain in our computation is set to $h=16$ to mimic the semi-infinite flow considered in \cite{Biau2016}. The lines show our computational results, while the three filled dots are extracted from Table 1 of \cite{Biau2016}. The $T_f$ in Biau \cite{Biau2016} needs to be interpreted as the elapsed time from $T_0$, rather than from $0$, as confirmed by Professor Biau (private communication). Here, $T_0$ denotes the starting time in the transient growth calculation, which will later be referred to as $t_0$ in our work. ($b$) Validation of the 2D instantaneous/momentary instability analyses against the results in \cite{Luo2010} and \cite{Blondeaux2021}. Their results were manually extracted from the respective papers. The red areas represent decelerating phases, and white areas represent accelerating phases.}
	\label{figureverification}
\end{figure}

When applying the FTLE to our linear time-periodic system, the deformation gradient of the linearised flow map reads 
\begin{align}
\nabla F^t_{t_0}(\boldsymbol {\tilde u}(t_0))  =\lim\limits_{\delta t \to 0}\displaystyle\prod_{j=1}^{n} e^{{\mathbf L}(t_j)\delta t}, 
\end{align}
following \cite{Farrell1996}. Here, ${\mathbf L}(t_j)$ is the linearised operator in Eq. \ref{lineareq}. Note that $\nabla F^t_{t_0}(\boldsymbol {\tilde u}(t_0))$ integrates the flow from $t_0$ to $t=t_0 +n\delta t$ with $t_j\in\big(t_0+(j-1)\delta t, t_0+j\delta t\big)$, and correspondingly, the time-ordering product $\Pi$ ensures that the dynamics propagates in the positive temporal direction. To quantify the growth or decay rate of the disturbance as $t\to\infty$, the first Lyapunov exponent \citep{Farrell1996} is defined as 
\begin{align}
 \Lambda_\infty =\displaystyle \lim_{t\to\infty}\sup \frac{\ln || \nabla F^t_{t_0}(\boldsymbol {\tilde u}(t_0))  ||_2}{2t} ,
\end{align} 
which implicitly assumes that the disturbance at $t_0$ has a unit norm.  A closely related concept is the transient growth, defined as the maximum energy amplification 
\begin{align}
 G(t_0, t)=\displaystyle\max_{{\mathbf u}(t_0)\ne \bf 0} \frac{\|\mathbf u(t)\|_2}{\|\mathbf u(t_0)\|_2} 
\end{align}
optimised over all possible initial conditions; see \cite{Schmid2001}. We will apply the transient growth analysis to the finite Stokes layer. To calculate the maximised energy amplification $G(t_0,t)$, we follow the direct-adjoint looping algorithm \citep{Luchini2014}, which has also been adopted by \cite{Biau2016}. Figure \ref{figureverification}($a$) validates our calculation in the case $h=16$ against those in the semi-infinite Stokes layer calculated by \cite{Biau2016}.

The Lyapunov exponent $\Lambda_\infty$ at large time ($t\rightarrow\infty$) is equivalent to the Floquet exponent in a time-periodic system. It concerns the flow behaviour at asymptotically large times. 
To investigate the intracyclic dynamics in the Stokes layer, we calculate the first FTLE \citep{Haller2015,Shadden2012} 
\begin{align}
\Lambda(t_0,t) = \sup_{\color{black}{\mathbf u}(t_0)\ne \bf 0} \frac{\ln\Big( \frac{\| \mathbf u(t)\|_2}{\| \mathbf u(t_0)\|_2} \Big)}{2(t-t_0)},
\end{align}
with the disturbance initiated at time $t_0$ and evolving until time $t$. The FTLE measures the maximum 'stretching' effect that a system can have in the time period $[t_0,t]$. Although it does not directly indicate the growth rate of a perturbation over time, it suggests the potential for maximum transient amplification within the system \citep{Kern2021}. 

\begin{table}
\centering
\begin{tabular}{c|cc}
\textrm{Cases}&
\textrm{\cite{Blennerhassett2006}}&
\textrm{Our results}\\ \hline
$Re=0.1, h=8$ &  -0.08833 (even) &  -0.08832989  \\
$Re=0.1, h=8$   &  -0.18181 (odd) &  -0.18180475 \\
$Re=570, h=16$ &  -0.06572 (even)& -0.06572140\\
$Re=570, h=16$ &  -0.11620 (odd)&  -0.11619931 \\
$Re=750, h=16$ &  -0.06695 (even)& -0.06694976\\
$Re=750, h=16$ &  -0.11951 (odd)&  -0.11951087\\
\end{tabular}
\caption{\label{table1}%
Validation of the 2-D Floquet analysis against the results in \cite{Blennerhassett2006} at $\alpha=0.3, \gamma=0$. Even and odd indicate the symmetry of the corresponding least-stable eigenfunction with respect to the channel centerline. We use $N_y=99$ and $N_f=170$ for the validation.
}
\end{table}

Existing theories on the instability of Stokes layers include Floquet theory and instantaneous/momentary stability theory, and will also be conducted in this work for comparison. Briefly, Floquet analysis assumes the solution ansatz 
\begin{align}
{{\bf u}}(x,y,z,t) = \displaystyle\Big[\sum_{n=-\infty}^\infty {{\bf u}}^{(n)}(y) e^{int}\Big] e^{i\alpha x + i\gamma z -i\lambda t}+ c.c.
\end{align} 
where $\lambda$ is the Floquet exponent encompassing the growth/decay rate and frequency of the disturbance, and $ c.c.$ indicates the complex conjugate of the preceding term. The infinite Fourier series needs to be truncated in numerical calculation, i.e., $n\in[-N_f,N_f]$. We consider a sufficient large $N_f$ ($>0.8\alpha Re$), following \cite{Thomas2011}. Another theory, the instantaneous/momentary stability theory \citep{Luo2010,Blondeaux2021}, assumes a solution form that reads ${{\bf u}}(x,y,z,t) =  \tilde{{\bf u}}(y) e^{i\alpha x + i\gamma z -iRe\int \lambda_2(t) dt}+ c.c.$ with $\lambda_2$ indicating the stability/instability of the disturbance. Verification of our Floquet analysis of the Stokes layers is shown in Table \ref{table1} and that of the instantaneous/momentary stability theories in figure \ref{figureverification}($b$).

\section{Results and discussion} \label{Results}


\subsection{General results}

\begin{figure}
	\centering
	\includegraphics[width=0.61\textwidth,trim= 0 65 160 5,clip]{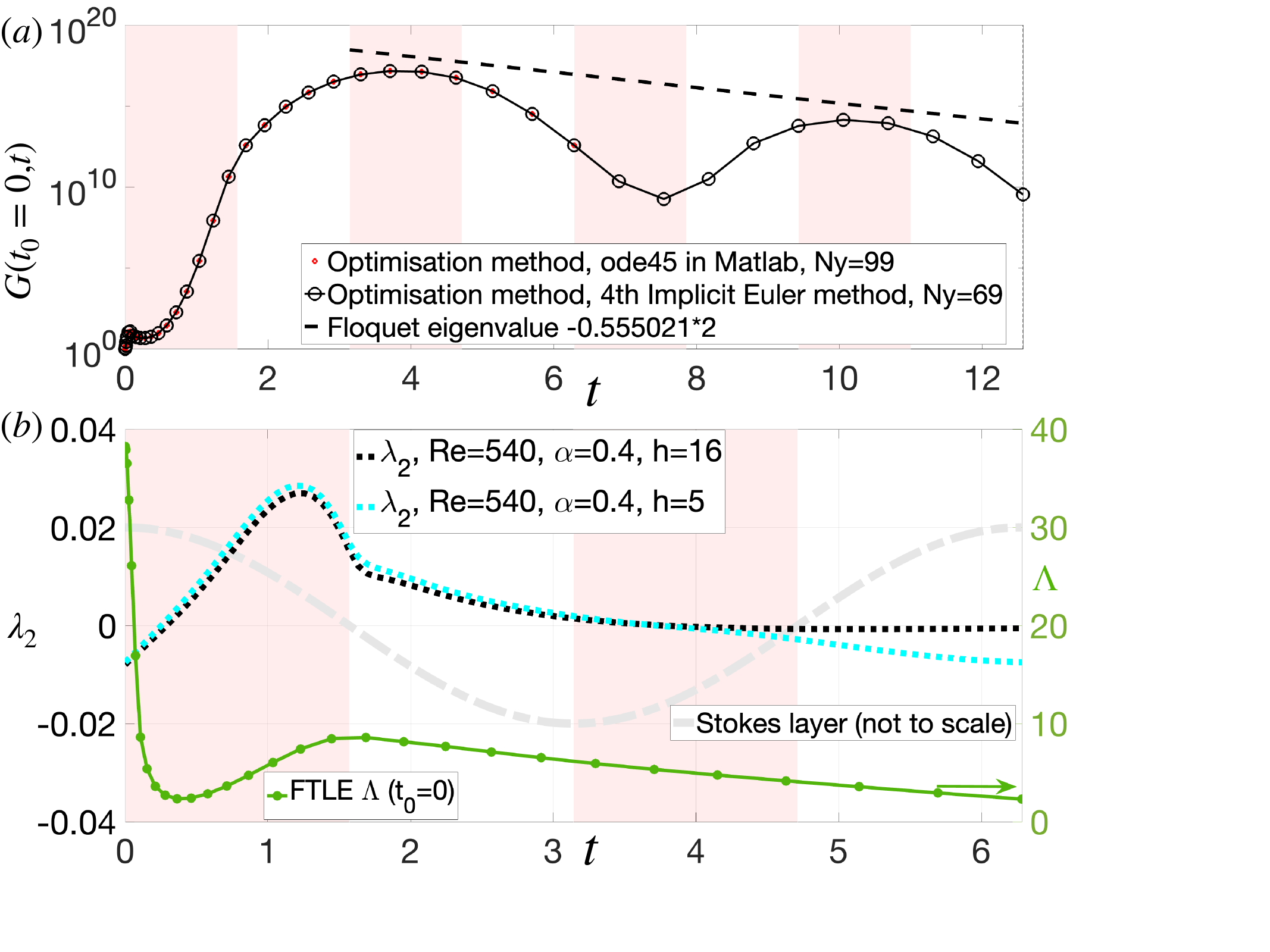}
\\	
	\includegraphics[width=0.48\textwidth,trim= 150 30 130 5,clip]{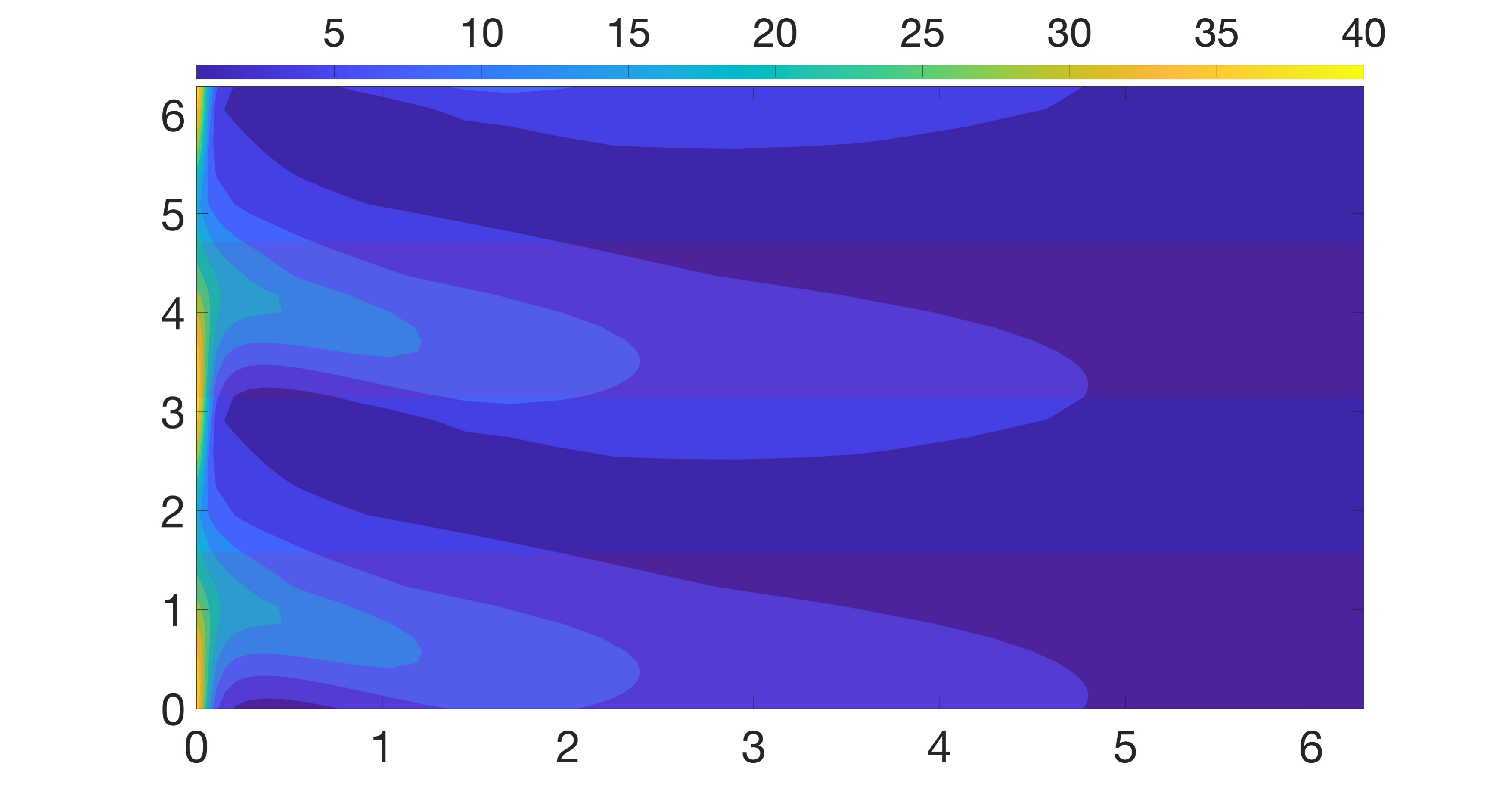}
	\put(-190,53){$t_0$}	
	\put(-100,-8){$\Delta t$}	
	\put(-185,115){$(c)$}		
	\includegraphics[width=0.48\textwidth,trim= 150 30 130 5,clip]{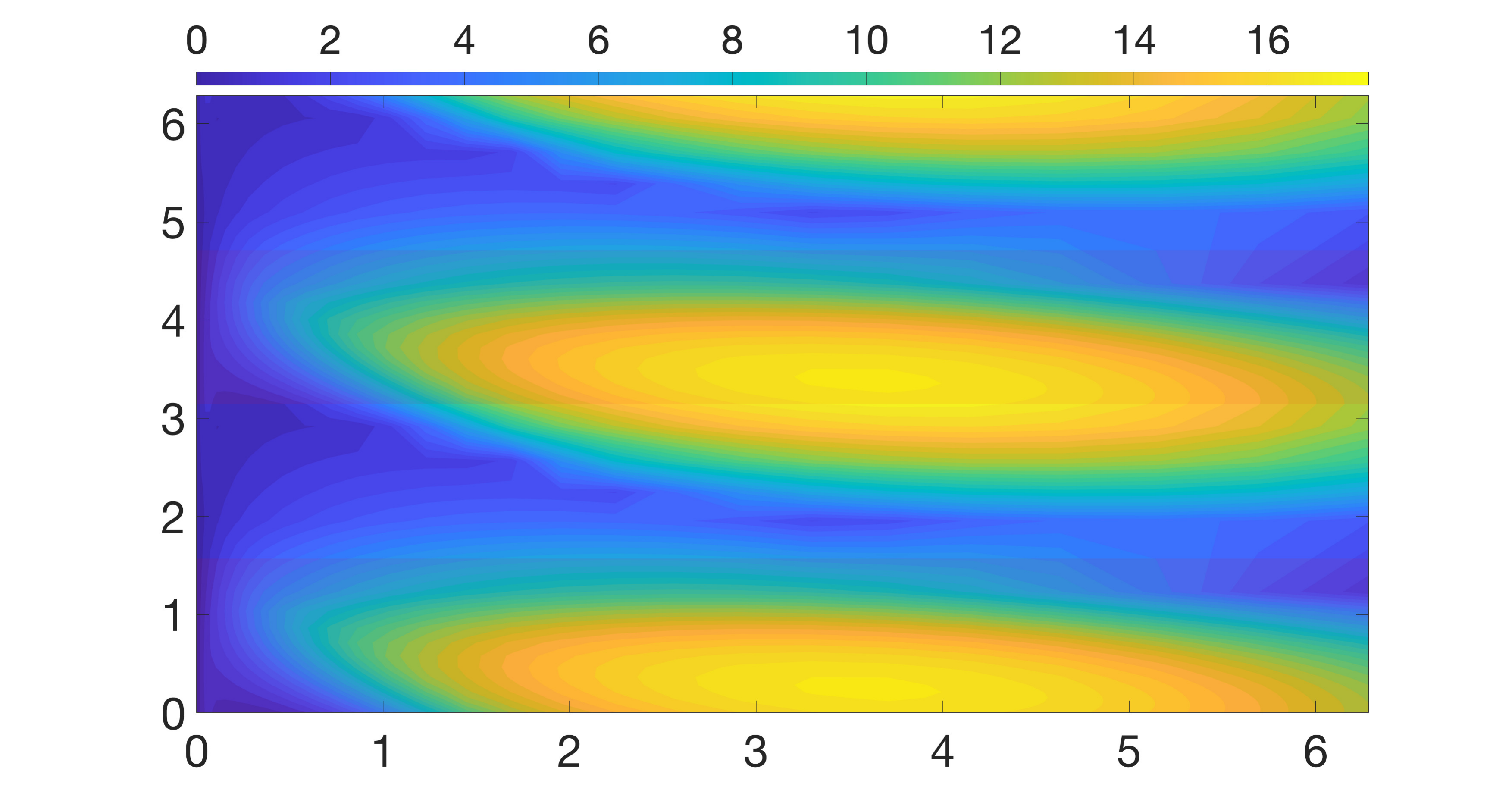}	
	\put(-100,-8){$\Delta t$}
	\put(-185,115){$(d)$}		
%
	\caption{Stability analyses of a typical 2-D finite Stokes layer with $Re=540, \alpha=0.4, h=5$. ($a$) Transient growth calculated using two time-integration methods (see the legend) and the Floquet decay rate at large time. ($b$) The growth rate $\lambda_2$  in the instantaneous/momentary stability analyses (left-hand $y$-axis) and the first FTLE $\Lambda$ (right-hand $y$-axis). ($c$) Distribution of FTLE as a function of the starting time $t_0$ and the integrated period $\Delta t\ (=t-t_0)$. (d) The corresponding transient growth $G(t_0,\Delta t)$ on a base-10 logarithmic scale. } 
	\label{figure1}
\end{figure}

\begin{figure}
	\centering
	\includegraphics[width=0.67\textwidth,trim= 60 15 70 0,clip]{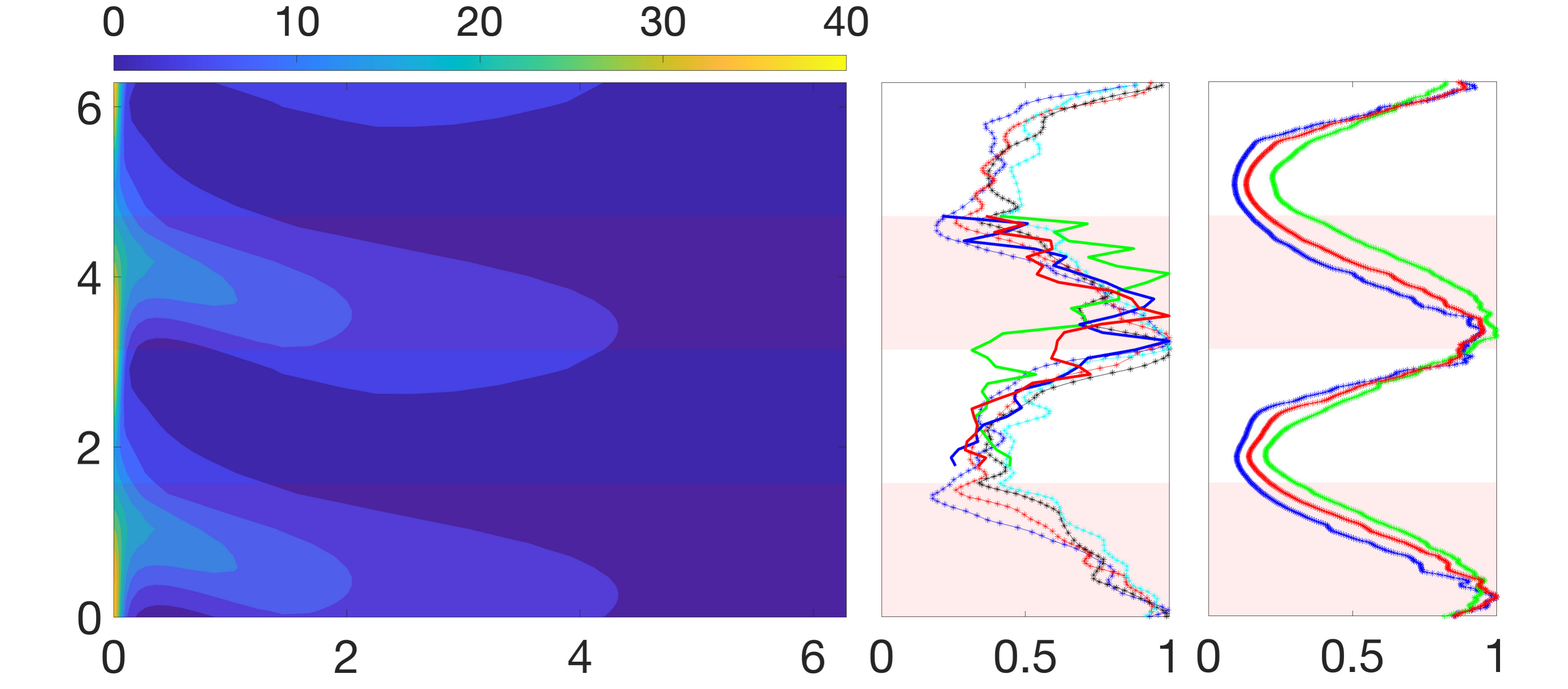}
	\put(-190,-10){\large$\Delta t$}
	\put(-265,53){\large$t_0$}		
	\put(-86,-10){\large$t$}	
	\put(-26,-10){\large$t$}				
	\put(-270,113){$(a)$}	
	\put(-110,113){$(b)$$({\overline{{u'}^2}})^{1/2}/max({\overline{{u'}^2}})^{1/2}$}	
	\put(-10,113){$(c)$}			
	\\
	\includegraphics[width=0.67\textwidth,trim= 0 300 0 0,clip]{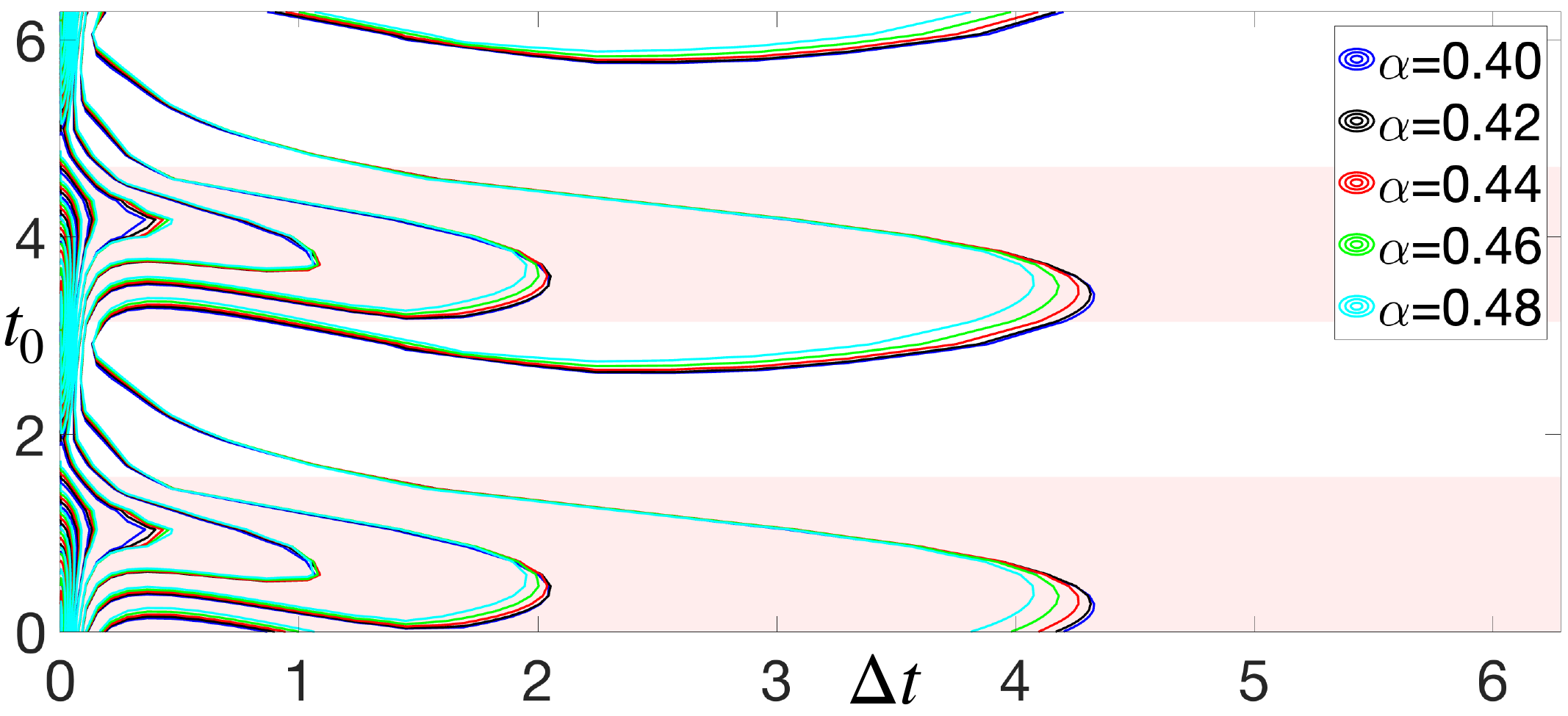}
	\put(-270,110){$(d)$}	
	\caption{($a$) Distribution of FTLE as function of the starting time $t_0$ and the integrated period $\Delta t\ (=t-t_0$). The parameters are $Re=540, \alpha=0.4, \gamma=0, h=10$. ($b$) Normalised axial turbulence intensity $({\overline{{u'}^2}})^{1/2}$ digitally extracted from the experimental literature. Lines with symbols from figure 9 of \cite{Akhavan1991}. \cite{Akhavan1991} concerns the Stokes layer in a pipe with $Re=540, h=10.6$ (or in their notations $Re^\delta=1080, \Lambda=10.6$) and their data are normalised by the maximum value in time at each radial location $r/R=0.992$ (blue), $0.95$ (red), $0.85$ (black) and $0.75$ (cyan), respectively, where $R$ is their pipe radius. Their experimental data have been shifted in time by $\pi/2$ to be consistent with our wall oscillation signal (i.e., $\cos t$). Thick lines without symbols from figure 11 of \cite{Hino1983}. \cite{Hino1983} studies the Stokes layer in a duct with $Re=438, h=12.8$ (or in their notations $R_\delta=876, \lambda=12.8$) and their data are normalised by the maximum value in time at each vertical location $0.01$ (blue), $0.05$ (red) and $0.1$ (green), respectively, where $d$ is their channel height. ($c$) Normalised axial turbulence intensity in our nonlinear simulation (to be detailed in Section \ref{nonlinear}). Starred blue curve is for $y=9.9818$; starred red curve is for $y=9.7773$; starred green curve is for $y=8.0545$.
	($d$) Effect of streamwise wavenumber $\alpha$ on the first FTLE $\Lambda$ distributed in the $t_0-\Delta t$ space. The other parameters are the same as those in $(a)$.}
	\label{figure2}
\end{figure}
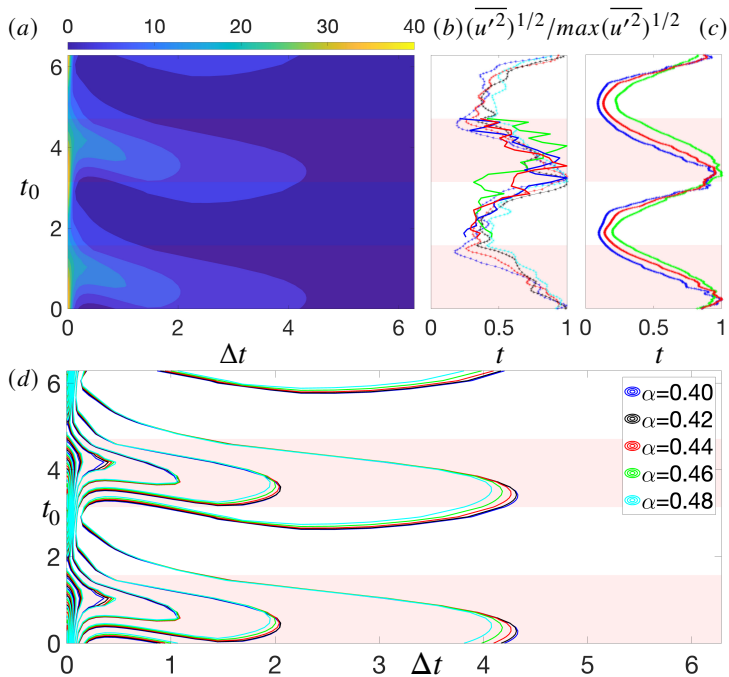

Figure \ref{figure1} shows the comparison of the stability analyses applied to the finite Stokes layer at $Re=540, \alpha=0.4, \gamma=0, h=5$. This $Re$ is close to values investigated in the experiments, e.g., by \cite{Akhavan1991} on the Stokes layer in a pipe and by \cite{Hino1983} on the Stokes layer in a duct (see the caption of figure \ref{figure2} for details). The streamwise wavenumber $\alpha=0.4$ corresponds closely to the wavelength of the most unstable disturbance, approximately 15, in the nonlinear simulations of \cite{Thomas2014} at $Re=600$, which employed the same length scale as ours. 

At these parameters, the flow is linearly stable according to the Floquet theory, as indicated by the dashed line in figure \ref{figure1}($a$). Figure \ref{figure1}($b$) shows that the instantaneous/momentary stability analysis of this flow predicts a negative growth rate $\lambda_2$ at the beginning of the first deceleration phase (red shade). For the chosen parameters, this contradicts the experimental observation that ``turbulence appeared explosively towards the end of the acceleration phase of the cycle and was sustained throughout the deceleration phase" \citep{Akhavan1991}.

Conversely, the transient growth $G(t_0=0,t)$ in figure \ref{figure1}($a$) initially shows a quick increase around $t=0$ and further climbs to $10^{17}$ in the second decelerating phase at around $t\approx4$. The substantial transient growth is consistent with the calculation by \cite{Biau2016} for the semi-infinite Stokes layer and the significant flow response to an impulse excitation studied by \cite{Thomas2014} (c.f. their figure 1). The mechanical explanation for this 2-D transient growth has been attributed to the Orr mechanism by \cite{Biau2016}. Biau also confirmed that three-dimensional (3-D) transient growth is less significant than its 2-D counterpart. Given that the maximum transient growth is exceptionally large, even small systematic or external perturbations may be amplified appreciably. This may explain why no experiments have reported carefully-controlled Stokes layers reaching a supercritical transition. In short, despite the Floquet stability of the Stokes layer, a subcritical transition likely occurs.

To investigate the intracyclic instability in the Stokes layer from the lens of the FTLE, the first FTLE $\Lambda$ is computed and shown in figure \ref{figure1}($b$) with $t_0=0$ as a function of $t$. The result for $\Lambda$ presents a different trend than the $\lambda_2$ in the instantaneous/momentary stability analysis and approaches the negative Floquet exponent $\lambda$ as $t\to\infty$ in this case. Figure ($d$) shows the corresponding $G(t_0,\Delta t)$ on a base-10 logarithmic scale. It can be seen that the global maximum transient growth is found at approximately $\Delta t\in[3,4]$ after the initial disturbance is imposed. Besides, the global maximum transient growth takes place in the decelerating phases (see the red areas), which is an encouraging result.

\subsection{The FTLE results compared to experiments}

The focus of the work is to test the comparison of the FTLE and transient growth with the experimentally observed intracyclic instability in the  Stokes layer flow. An appealing comparison of the FTLE with the experimental data is shown in figure \ref{figure2}($a$), which presents the distribution of $\Lambda$ as a function $t_0$ and $\Delta t\ (= t-t_0)$. We now take $h=10$ to be consistent with the experimental set-up (see the caption for details).
Overall, the distribution implies relatively larger growth rates during the decelerating phases (red shades in the contour figure), indicating stronger flow instability, which may lead to more intensive turbulence. This appears to be consistent with the experimental observation that strong turbulent activity bursts and persists in the decelerating phases, but weaker turbulence occurs during the accelerating phase; see figure 9 in \cite{Akhavan1991} for the turbulence intensity, figure 6 in \cite{Hino1976} for the velocity variation and figures 15 and16 in \cite{Hino1983} for the turbulence energy production. As pointed out by one of the reviewers, the experiments were conducted with oscillating pressure gradient and stationary walls. Indeed, conducting experiments with physically oscillating walls is not practically feasible. However, from a mathematical standpoint, the Navier–Stokes equations with oscillating wall boundary conditions are equivalent to those with an oscillating pressure gradient and stationary walls. Therefore, it is reasonable to compare our results directly with the existing experiments involving oscillatory pressure-driven flows.

To facilitate the comparison, the experimental results of axial turbulence intensity extracted from \cite{Akhavan1991,Hino1983} are reproduced in figure \ref{figure2}$(b)$ at various radial/vertical locations in their experiments. 
Notably, the ``eggplant" regions within $\Delta t \in [1, 2\pi]$ in our $\Lambda$ distribution, which encompass the maximum transient growth, precisely align with the peak positions of axial turbulence intensity observed in the experiments. Figure~\ref{figure2}($c$) shows the results of our nonlinear simulations, which will be discussed in detail in Section~\ref{nonlinear}. As a brief remark, the intracyclic instability observed in the nonlinear simulations is also consistent with the FTLE results. In short, figures \ref{figure2}($a,b$) reveal a strong correlation between the distribution of the FTLE and the experimentally-observed axial turbulence intensity, providing convincing evidence of intracyclic dynamics in the finite Stokes layer. Together with \cite{Biau2016}, these results imply a linear energy amplification mechanism underpinning the turbulence generation cycle in transitional Stokes layers. 

The above result focuses on the Stokes layer for a single wavelength. To assess the robustness of the FTLE distribution across different wavelengths, figure \ref{figure2}($d$) presents line contours of the FTLE for $\alpha=0.40:0.02:0.48$. This range of $\alpha$ encompasses the most significant transient growth in this subcritical Stokes layer, see also figure \ref{figurealpRe} below. The same FTLE distribution is observed across all these wavenumbers, demonstrating that the energy amplification mechanism in the finite Stokes layer extends over a spectrum of wavenumbers linked to the most amplified disturbances.




\begin{figure}
	\centering
	\includegraphics[width=0.9\textwidth,trim= 30 185 60 5,clip]{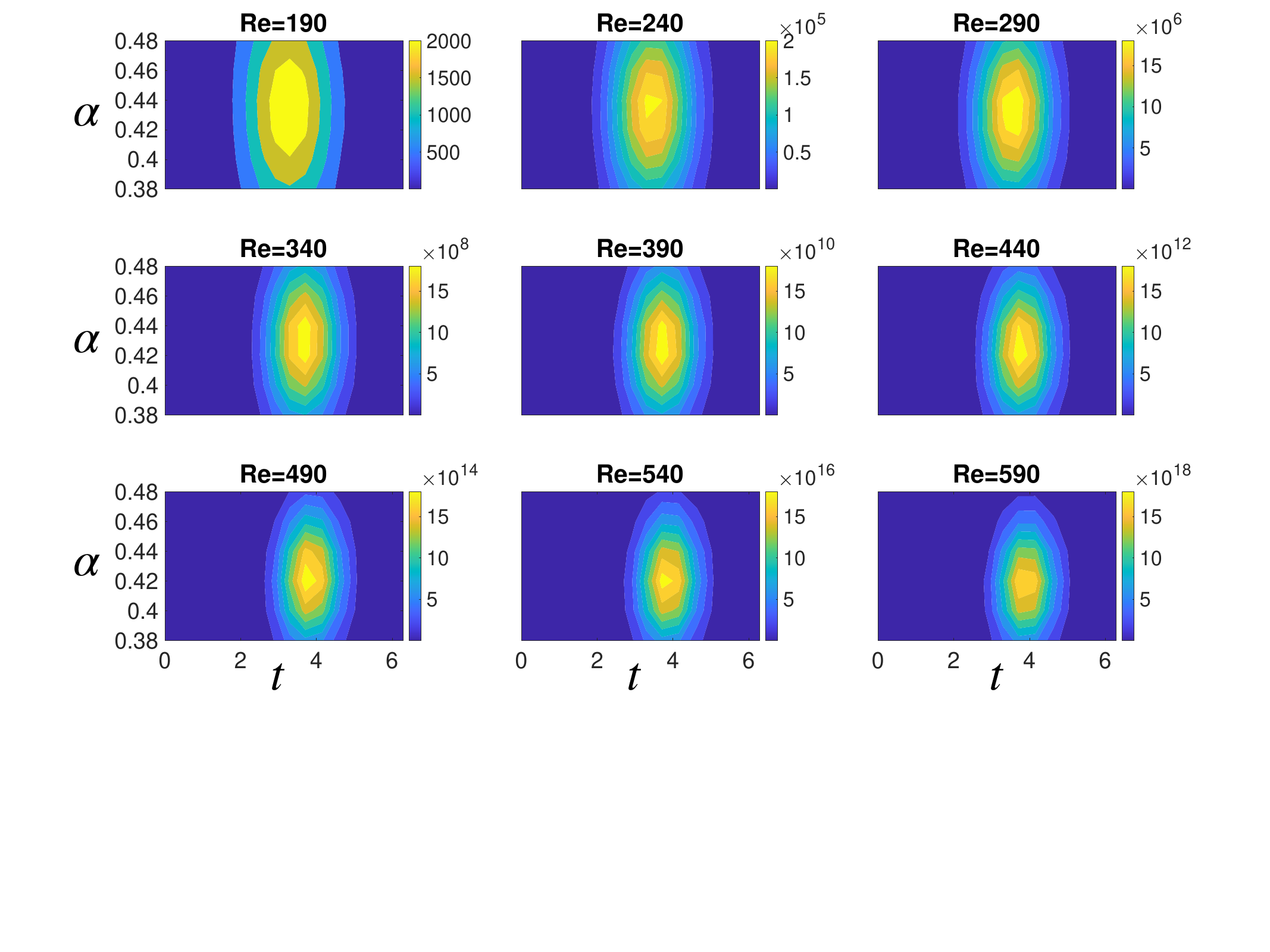}
	\caption{Effect of $\alpha$ and $Re$ on the transient growth $G(t_0=0,t)$ (color contour) in 2D finite Stokes layers with $h=5$.  }		
	\label{figurealpRe}
\end{figure}

\subsection{Parametric study: Effects of $Re$, $h$ and $\omega$}

 A parametric study of the effects of $Re$ and $\alpha$ on the transient growth $G(t_0=0,t)$ in the 2-D finite Stokes layer is shown in figure \ref{figurealpRe}. It is evident that transient growth rises with $Re$ within this subcritical range. This is consistent with the calculation by \cite{Biau2016}, who similarly observed the monotonic increases of the transient growth in the subcritical semi-infinite Stokes layer. The optimal wavenumber for transient growth is found to be approximately $\alpha = 0.42$, thereby supporting our choice of $\alpha = 0.4$ in the present investigation.

\begin{figure}
	\centering
	\includegraphics[width=0.6\textwidth,trim= 0 0 210 0,clip]{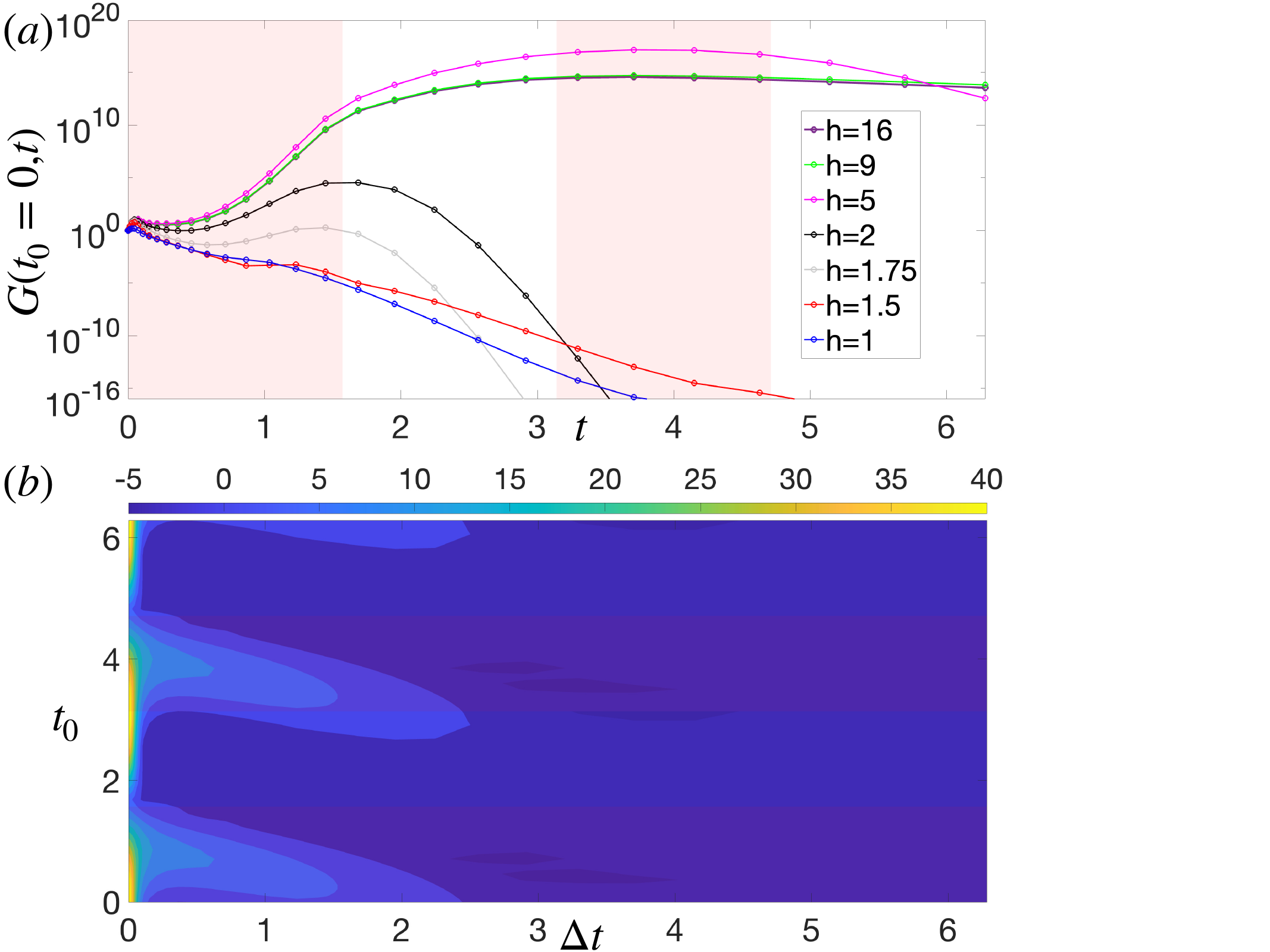}	
	\caption{($a$) Effect of nondimensional channel half-height $h$ on the transient growth $G(t_0=0,t)$. ($b$) Distribution of the first FTLE $\Lambda$ in the $t_0-\Delta t$ plane for the case $h=2$. The other parameters are $Re=540, \alpha=0.4, \gamma=0$. }
	\label{figure4}
\end{figure}

Next, we explore the effect of the half-wall distance $h$ on the FTLE in the finite Stokes layers. Large-$h$ flows have been widely studied in the literature, while small-$h$ cases, though less explored, are relevant in fields such as biology and microrheology \citep{Mitran2008}. Figure \ref{figure4}($a$) shows the 2-D transient growth at $Re=540, \alpha=0.4$ for various $h$. For relatively large $h$ ($h\ge9$), the flow resembles its semi-infinite counterpart, presenting significant transient growth. The effect of changing $h$ is minor in this range. The maximum $G(t_0=0,t)$ is found at $h=5$; see the magenta line. Upon reducing $h$, the transient growth decreases drastically. This suggests that as $h$ decreases, the linear amplification mechanism weakens. This trend is consistent with the observation by \cite{Hino1976} that ``The critical Reynolds number of the first transition decreases as the Stokes parameter increases" in the small-$h$ range. Their Stokes parameter is identified as $h$ in our work. To illustrate the feature of the $\Lambda$ distribution in the small-$h$ Stokes layers, we take the case $h = 2$ as an example, depicted in figure \ref{figure4}($b$). By comparing this result with that for $h=10$ in figure \ref{figure2} and that for $h=5$ in figure \ref{figure1} (corresponding to the maximum FTLE over all $h$'s), it is evident that the FTLE for $h=2$ is significantly smaller and exhibits a weaker history effect along the $\Delta t$-axis. Thus, our FTLE analysis suggests a weaker energy amplification mechanism in the small-$h$ Stokes layers. Based on this observation, we hypothesise that the small-$h$ Stokes layers, if they can transition to turbulence, will show a smaller temporal standard deviation in the normalised axial turbulence intensity, compared to the large-$h$ flows. We encourage future experimental research to confirm this finding by changing the dimensional channel half-height $d$.

In experiments involving the Stokes layer, nevertheless, the most practical way to vary the system parameters is by adjusting the wall-oscillation frequency $\omega$. Under our non-dimensionalisation scheme, changes in $\omega$ induce corresponding changes in $Re$, $h$ and $\alpha$. To investigate how transient growth varies with frequency, we consider a reference parameter set defined by 
\begin{align}
   Re_{\text{ref}} = \frac{U_0}{\sqrt{2\nu \omega_{\text{ref}}}} = 540\ \ \ \  \text{and}\ \ \ \ \ h_{\text{ref}} = \frac{d}{\sqrt{2\nu / \omega_{\text{ref}}}} = 3, 
\end{align}
while fixing $\alpha_{\text{ref}} = 0.4$. We choose to fix $\alpha = 0.4$ because, in realistic experiments, a spectrum of waves is typically excited rather than a single mode. Our parametric study (see figure~\ref{figurealpRe}) indicates that the maximum transient growth occurs near $\alpha \approx 0.4$, justifying its use to focus on the most dangerous scenario. Defining the frequency ratio as $r = \omega / \omega_{\text{ref}}$, the new dimensionless parameters become 
\begin{align}
   Re = \frac{Re_{\text{ref}}}{\sqrt{r}}, \quad h = h_{\text{ref}} \sqrt{r}    
\end{align}
as functions of $r$.

\begin{figure}
	\centering
	\includegraphics[width=0.65\textwidth,trim= 40 30 40 20,clip]{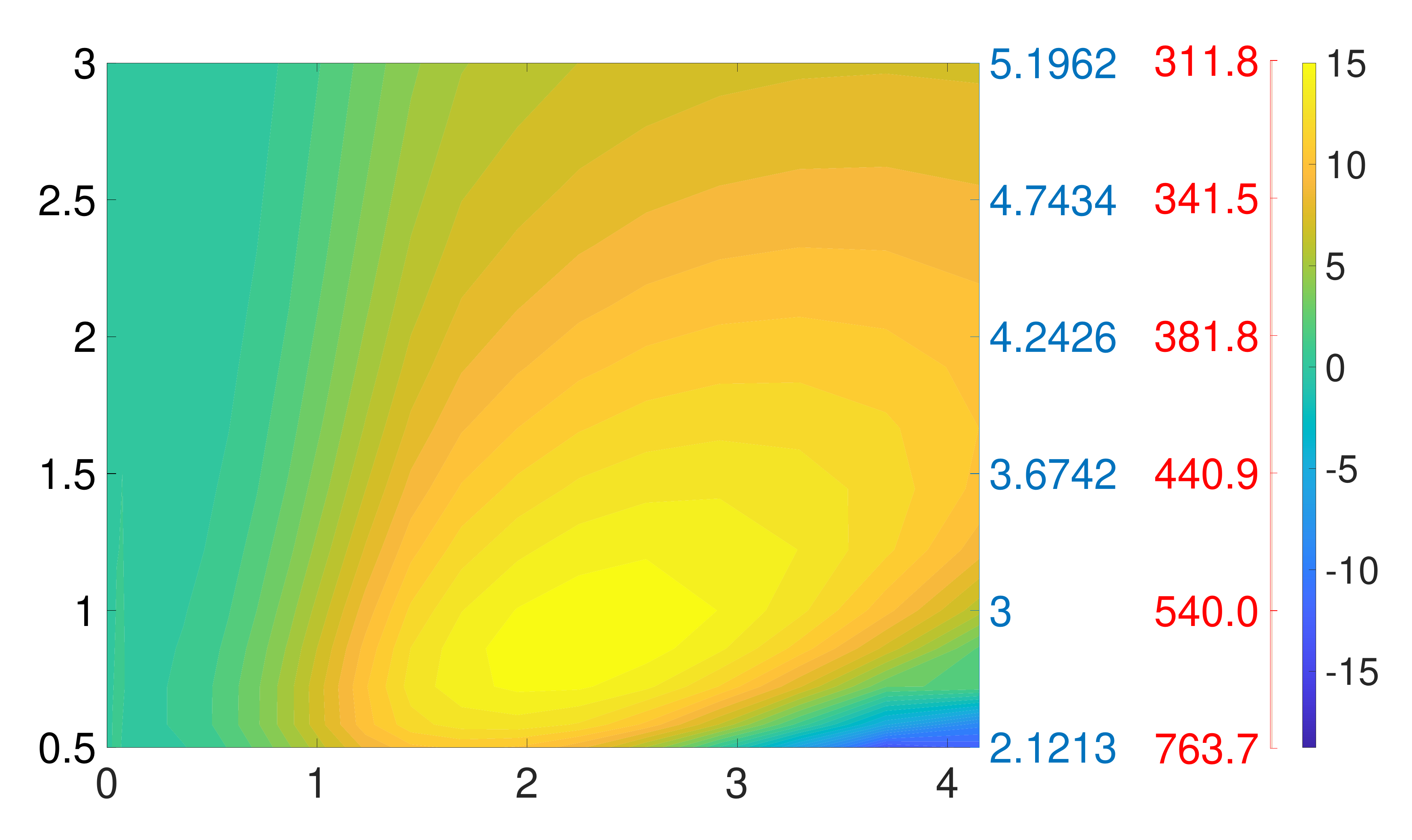}
	\put(-280,70){$r=\frac{\omega}{\omega_{ref}}$}	
	\put(-140,-5){$t$}	
	\put(-36,150){$Re$}		
	\put(-60,150){$h$}			
	\caption{Contour plot of $G(t_0 = 0, t)$ on a base-10 logarithmic scale. The left-hand $y$-axis indicates the ratio $\omega / \omega_{\text{ref}}$, where $\omega_{\text{ref}}$ corresponds to the reference parameter set $(Re_{\text{ref}} = 540, h_{\text{ref}} = 3)$. The wavenumber is fixed at $\alpha = 0.4$ in all cases to capture the most amplified transient growth. The corresponding values of $h$ and $Re$ are also shown in blue and red, respectively.}
	\label{figure_omega}
\end{figure}
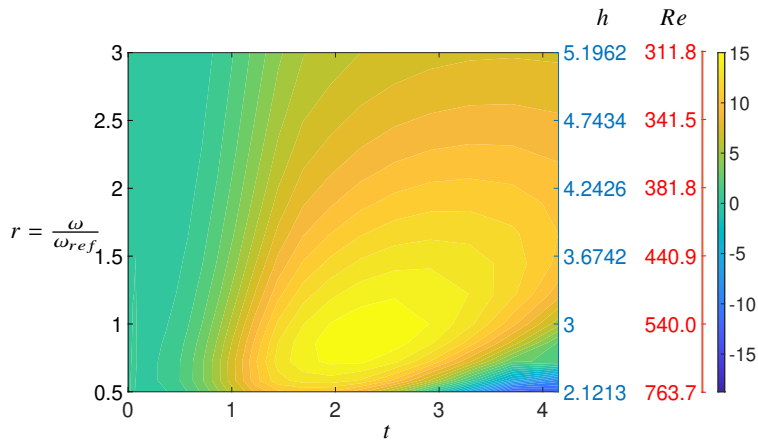

Figure \ref{figure_omega} implies that increasing $\omega$ (or equivalently $r$) leads to a non-monotonic response in the maximum transient growth with the initial disturbance imposed at $t_0=0$. From $r=0.5$ ($h = 2.1213, Re=763.7$) to $r=1$ ($h = 3, Re=540$), the maximum transient growth increases, indicating a destabilizing effect of increasing frequency. Beyond this point, however, further increases in frequency suppress transient growth. We compare this trend to those in the literature. \cite{Merkli1975} showed in their figure 5 that the maximum velocity in the flow increases with increasing frequency, suggesting a destabilizing effect associated with higher oscillation frequencies. The discrepancy between our results and those of \cite{Merkli1975} may arise from differences in the parameter regimes explored. As presented in our figure \ref{figure4} above, the value of $h$ seems to be playing an important role in determining the transient growth; however, \cite{Merkli1975} did not report the values of $h$, making a direct comparison difficult. 

To gain further insight, we refer to the experimental work of \cite{Hino1976}, where a comparable parameter, $\lambda$ (the same definition as our $h$), increases from 2.76 to 3.90 as the oscillation period $T$ is reduced from 6.0 s to 3.0 s,  effectively doubling the frequency. Table 1 of \cite{Hino1976} shows that the increase in $\lambda$ is accompanied by a transition from disturbed laminar flow (denoted by empty circles) to weak turbulence (denoted by filled circles), also supporting the destabilizing effect of increasing frequency. Notably, this transition in experiments occurs within the range $h \in [2.76, 3.90]$, which approximately aligns with the low-$h$ regime in our study, where a similar destabilizing trend is observed.

\subsection{Nonlinear evolution}\label{nonlinear}
The discussions above centre around the linear dynamics. To further extend the implication of the linear results, especially the intracyclic instability shown in figure \ref{figure2}, nonlinear simulations were conducted. The numerical method adopted is the classic pseudo-spectral method, and the details are presented in Appendix \ref{Appendix_numerical}. The computation domain, with the channel half-height as the reference length, is ($2\pi/0.4, 5,  2\pi$) discretised with a resolution $129\times 105\times 89$ along the $x,y,z$ directions. A constant time step $dt= 2\pi/250000\approx2.5132\times 10^{-5}$ is considered. The initial condition of the nonlinear simulation consists of the laminar Stokes layer solution at $t=0$ plus the optimal initial condition (whose kinetic energy is $10^{-10}$) targeting the maximum transient growth at $t=1$. A 3-D perturbation is additionally imposed to trigger the transition, following \cite{Biau2016}.

\begin{figure}
	\centering
	\includegraphics[width=0.99\textwidth,trim= 100 0 140 0,clip]{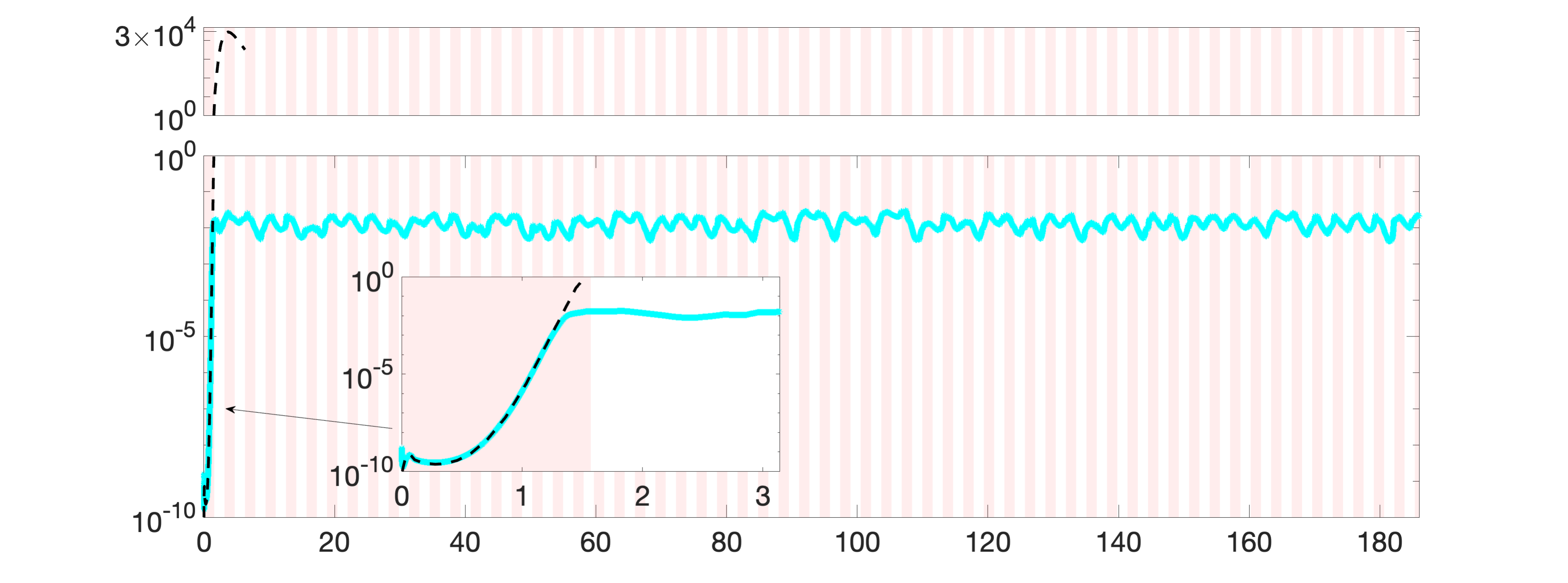}	
	\put(-185,-2){$t$}
	\put(-380,40){\rotatebox{90}{Perturbation energy $E'_k$}}
	\caption{Nonlinear evolution of the perturbation kinetic energy $E'_k(t) = \frac{1}{2h}\int ||{\bf{u}}(y,t) - {\bf{U}}_b(y,t) ||^2 dy$ (cyan line) at $Re=540, h=10$. Note that ${\bf{u}}(y,t)$ have been averaged over the $xz$ plane before the integration and ${\bf{U}}_b(y,t)$ is homogeneous in the $x$$z$ plane by definition. The dashed black line represents the envelope of the maximum transient growth. The total simulation time is approximately 186 time units, corresponding to 29.6 complete periods.  }
	\label{figure6}
\end{figure}

Figure~\ref{figure6} presents the evolution of the perturbation energy \( E'_k \), computed by integrating the kinetic energy of the velocity deviation from the laminar Stokes layer solution ${\bf{U}}_b(y,t)$ over the entire domain at \( Re = 540 \). The inset highlights that the evolution of \( E'_k \) closely follows the optimal growth envelope (calculated in the transient growth analysis), indicated by the black dot-dashed line; this is an outcome of the optimal initial condition as part of the initial condition used in our nonlinear simulation. The sharp energy rise at \( t = 0 \) results from the 3-D disturbance, which decays rapidly. After the initial transient, the flow reaches a statistically saturated state before the end of the first decelerating phase. It is evident that the perturbation energy consistently troughs during the accelerating phases (white background) and peaks during the decelerating phases (red shading). This behaviour aligns with both the FTLE analysis and the experimental observations shown in figures~\ref{figure2}($a,b$). To enable a more quantitative comparison, we also compute the phase-averaged axial turbulence intensity, \( ({\overline{u'^2}})^{1/2} \), normalised by its maximum value, at three different vertical positions ($y=9.9818, 9.7773, 8.0545$) approximately corresponding to the three vertical positions in \cite{Hino1983}; see the starred curves in figure~\ref{figure2}($c$). The axial turbulence intensity is calculated by subtracting the axial component of the laminar Stokes layer ${{U}}_b(y,t)$ from the raw data. The result in figure~\ref{figure2}($c$) confirms that the phase evolution of axial turbulence intensity relative to the laminar reference flow qualitatively captures the intracyclic instability of the perturbations.

\begin{figure}
	\centering
	\includegraphics[width=0.85\textwidth,trim= 70 0 60 0,clip]{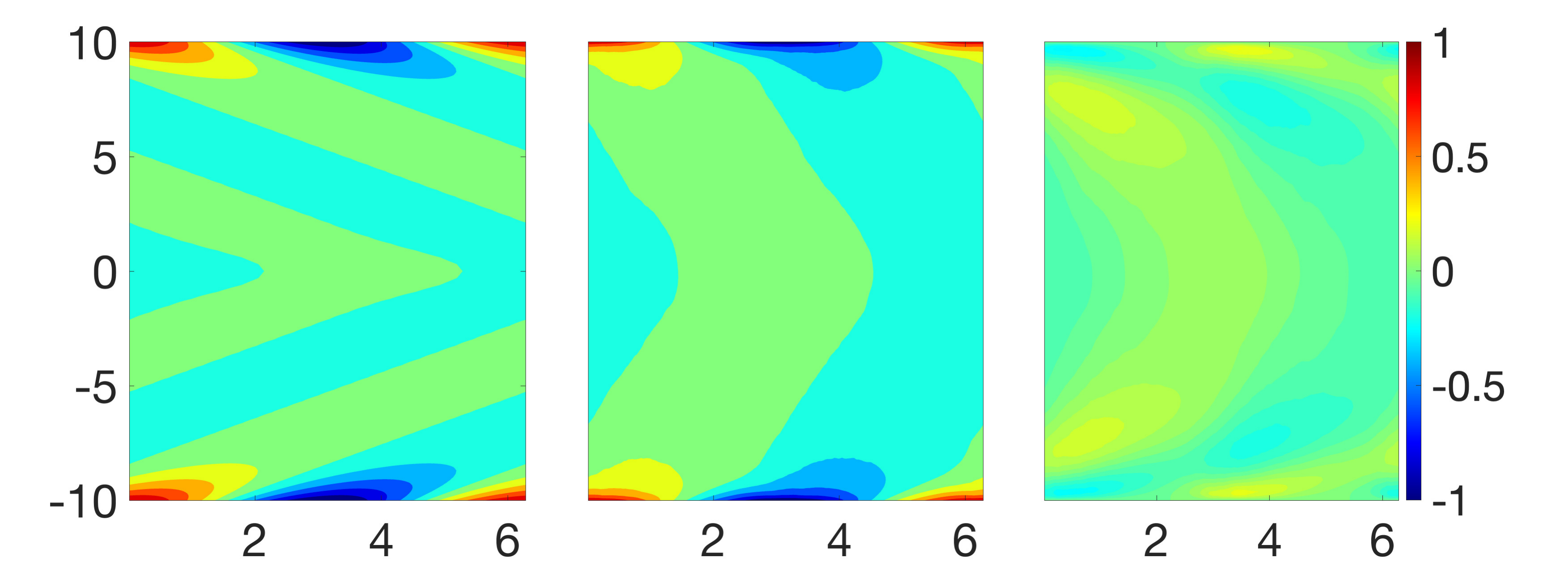}	
	\put(-333,128){\large $(a)$}
	\put(-213,128){\large $(b)$}	
	\put(-108,128){\large $(c)$}	
	\put(-333,63){\large $y$}
	\put(-153,-10){\large $t$}	
	\caption{Profiles of $(a)$ the laminar Stokes layer ${{U}}_b(y,t)$, $(b)$ the phase-averaged flow along the $x$ direction over 27 periods ${{\bar u}}(y,t)$, and $(c)$ the difference between the two, ${{U}}_b(y,t)-{{\bar u}}(y,t)$, at $Re=540, h=10$. Here, ${{\bar u}}(y,t)$ has been spatially averaged in the $x-z$ plane.}
	\label{figure7}
\end{figure}

As nonlinear effects distort the laminar base flow, it is instructive to compare the laminar Stokes layer ${\bf{U}}_b(y,t)$ with the phase-averaged flow, denoted as ${\bf{\bar u}}(y,t)$ averaged over the $xz$ plane. The $x$ component of the two flows is shown in Figure~\ref{figure7}. The nonlinear simulation was run for 186 time units, corresponding to 29.6 oscillation periods. To eliminate initial transients, the first two periods are discarded, leaving the remaining 27 periods to be used for the phase-averaging. This duration is considered sufficiently long, comparable with simulation times reported in studies such as 2$\sim$13 cycles in \cite{Vittori1998}, 50 cycles in \cite{Manna2015} and 10 cycles in \cite{Ebadi2019}. Figure \ref{figure7}($a$) displays the laminar Stokes layer in a $y$–$t$ diagram, and figure \ref{figure7}($b$) shows the phase-averaged flow. Although the near-wall velocity patterns appear similar in the two flows, the less tilted green/cyan stripes in the ${{\bar u}}(y,t)$ field suggest enhanced temporal coherence across the velocity field. The small difference, as shown in figure \ref{figure7}($c$), indicates the relative resemblance of the two flows.

\begin{figure}
	\centering
	\includegraphics[width=0.7\textwidth,trim= 70 0 60 0,clip]{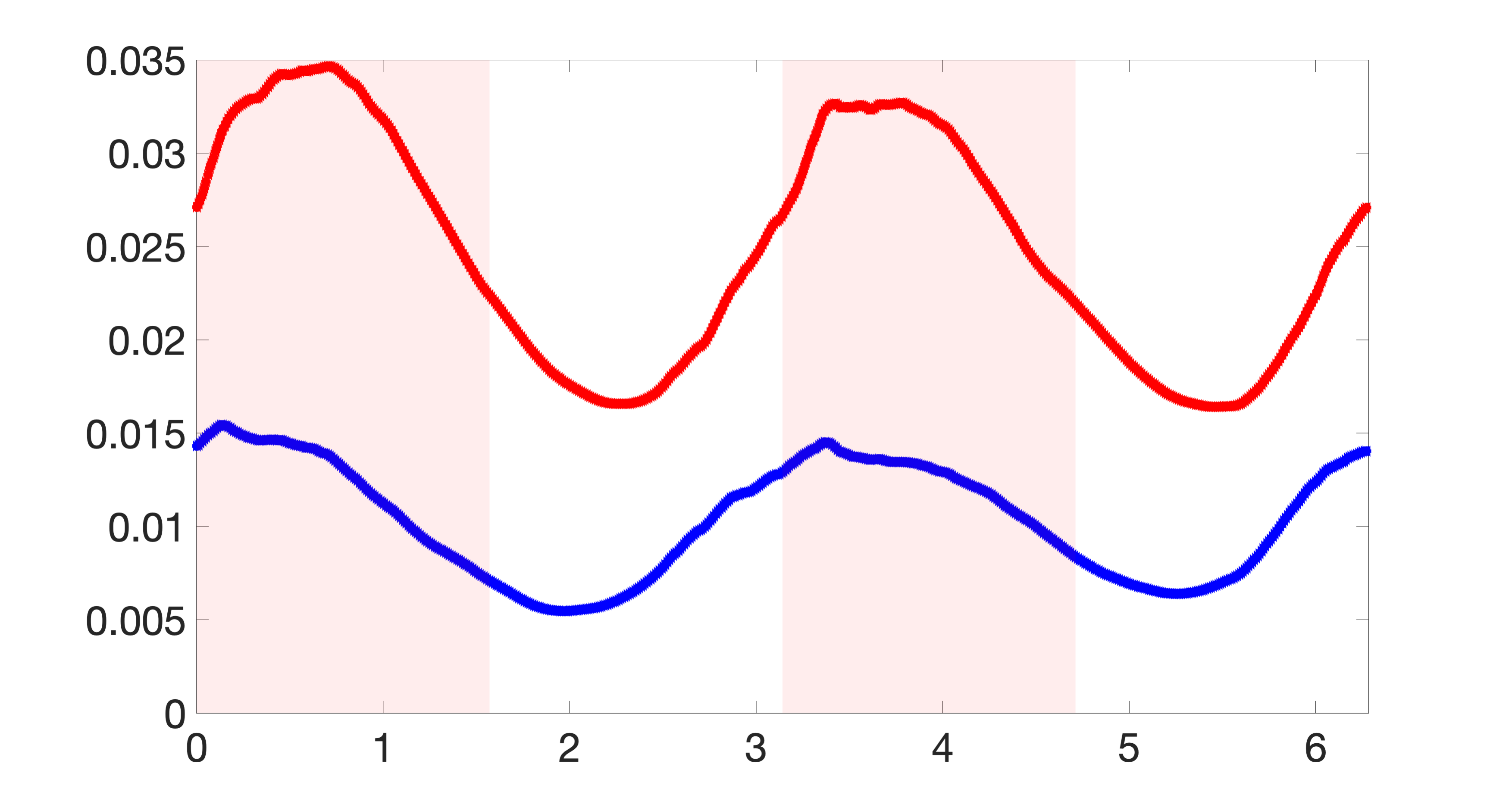}	
	\put(-283,75){\large $E'_x$}
	\put(-133,-0){\large $t$}	
	\caption{Phase-averaged $x$-component perturbation kinetic energy $E'_x(t)$ calculated with respect to the laminar flow (red, $E'_x(t) = \frac{1}{2h}\int ||{{u}}(y,t) - {{U}}_b(y,t) ||^2 dy$) and the time-mean flow (blue, $E'_x(t) = \frac{1}{2h}\int ||{{u}}(y,t) - {{\bar u}}(y,t) ||^2 dy$) at $Re=540, h=10$. Here, ${{\bar u}}(y,t)$ denotes the $x$ component of the phase-averaged flow and has been averaged over the $xz$ plane.}
	\label{figure8}
\end{figure}

To further compare the laminar base flow ${\bf{U}}_b(y,t)$ and the phase-averaged flow ${\bf{\bar u}}(y,t)$, figure~\ref{figure8} shows the $x$-component of the perturbation kinetic energy \( E'_x \) computed with respect to them. As observed, \( E'_x \) is smaller for the time-mean flow (blue) than for the laminar flow (red), suggesting that the nonlinear flow evolves towards a periodic state that is statistically closer to the phase-averaged flow, which is consistent with expectations. Besides, both curves exhibit the same intracyclic pattern observed in the experiments and the FTLE, demonstrating qualitative agreement with our results in figure \ref{figure2}. 

Following the evaluation of the two base flows, we note that obtaining the time-mean flow requires a computationally expensive nonlinear simulation, whereas the laminar solution can be derived analytically. Given their comparable performance in capturing key dynamical features, it may be worthwhile to explore modelling strategies for the Stokes layer based on both formulations and quantify the difference in their performance. In particular, a comparative resolvent analysis using the laminar base flow versus the phase-averaged flow could offer valuable insights into their respective effectiveness in capturing the flow’s response characteristics.

\section{Conclusion}
The flow transition in the Stokes layer is studied numerically by calculating its first FTLE in the linear regime. The distribution of the FTLE closely matches the intracyclic dynamics of the axial turbulence intensity observed in experimental studies of Stokes layers in channel and pipe, a phenomenon that earlier stability analyses were unable to capture accurately. Our nonlinear simulations also confirm the intracyclic instability, particularly the elevated disturbance amplification during the decelerating phase and the quenched flow instability during the accelerating phase, similar to the FTLE results. The underlying energy amplification mechanism is the non-normal transient growth in the time-dependent shear flow \citep{Farrell1996}, first revealed by \cite{Biau2016} for the semi-infinite Stokes layer. 

According to our calculation, this non-normal mechanism can significantly amplify systematic or external disturbances within the first oscillation cycle, leading to intracyclic instability despite the long-term stability predicted by the Floquet theory. This may explain the consistent experimental observations of subcritical transition in the confined Stokes layers. The agreement between the intracyclic instability observed in experiments and that predicted by the FTLE analysis underscores the significance of non-normality in flow transition, which is the main contribution of this investigation. 

In conventional rectilinear wall-bounded shear flows, the non-normal mechanism has been widely studied as a key factor in the energy amplification mechanism \citep{Schmid2001} and turbulence generation process; see \cite{Jimenez2013, Lozano-Duran2021} among many others. In periodically-driven time-dependent shear flows, such as the Stokes layer and other pressure-driven flows \citep{Pier2021,Kern2021}, the non-normal mechanism plays an equally significant role in flow transition. In brief, this investigation enhances our understanding of turbulence generation in the Stokes layer and may help to provide new insights into other time-dependent systems, including climate dynamics and biological flows.

This work opens several avenues for future research to deepen our understanding of flow transition in Stokes layers and other time-dependent flows. First, the FTLE analysis of the direct numerical simulation results should be conducted to account for the nonlinear Stokes layers. Nonlinear FTLE, computed without linearization, not only depends on $t_0$ and $\Delta t$, but also varies with spatial location. Since experimental results can be viewed as nonlinear realisations, we anticipate that the distribution of the FTLE in the nonlinear flow may resemble the linear results presented here. Second, flow control can be more effectively applied to the Stokes layer by targeting the identified energy amplification mechanism. Finally, future experiments should detail the influence of channel heights on the turbulence generation in finite Stokes layers, where the reduced transient growth will lead to a more stable flow and alter the turbulence dynamics. Such experiments would provide validation of the linear mechanisms identified in this study and explore the subsequent nonlinear dynamics. Further quantification of the differences between the laminar Stokes layer and the phase-averaged flow, particularly in the context of flow modelling, is also a worthwhile direction for future investigation.

Declaration of Interests. The author reports no conflict of interest.

This work is supported by Ministry of Education, Singapore via the grant WBS no. A-8001172-00-00. We sincerely thank Professor D. Biau for his kind assistance in helping us to verify our transient growth calculations.

\begin{appendix}
\section{Numerical methods for direct numerical simulations}\label{Appendix_numerical}
The numerical algorithm for the in-house direct numerical simulations (DNS) code used in Section \ref{nonlinear} is explained in this appendix. A Fourier-Chebyshev-Fourier spatial discretisation scheme is employed along the $x,y,z$ directions, respectively. The incompressible Navier-Stokes equations (\ref{eq21}) are time-advanced using a Crank-Nicolson Adams-Bashforth (CNAB) scheme
\begin{align}
\frac{\tilde{\bf  u}^{n+1}-\tilde{\bf u}^n}{\Delta t} + \frac{3}{2} Re \tilde{\bf N}^n - \frac{1}{2} Re \tilde{\bf N}^{n-1}&= -\frac{Re}{2}\nabla \tilde p^{n+1} -\frac{Re}{2}\nabla \tilde p^n  + \frac{1}{2} \frac{\nabla^2 \tilde{\bf u}^{n+1}}{2} + \frac{1}{2}\frac{\nabla^2 \tilde{\bf u}^{n}}{2}.    \label{CNABscheme}
\end{align}
where the convective term $\tilde{\bf N}=(\tilde{\bf u}\cdot \nabla )\tilde {\bf u}$ is expressed using the rotational formulation and the conservative formulation alternately at each time step. The CNAB method has been proven to be effective for simulating low-$Re$ flows \citep{Kim1987}. The above numerical equation is solved together with the continuity condition with the imposed boundary conditions $\tilde {\bf u}^{n+1}(\pm h)=\mathbf U_b(t^{n+1})$ at the moving walls. This is achieved following the influence matrix method introduced in \cite{Madabhushi1993}. After some manipulation, Eq. (\ref{CNABscheme}) becomes
\begin{align}
2\frac{\tilde{\bf  u}^{n+1}}{Re\Delta t} -  \frac{1}{2Re} \nabla^2 \tilde{\bf u}^{n+1} + \nabla \tilde p^{n+1} =\tilde{\bf r}^n =\frac{2\tilde{\bf u}^n}{Re\Delta t}  - 3 \tilde{\bf N}^n +  \tilde{\bf N}^{n-1}  - \nabla \tilde p^n  + \frac{1}{2Re} \nabla^2 \tilde{\bf u}^{n}.
\end{align}
According to \cite{Madabhushi1993}, discretisation error must be accounted for in the numerical equation. Denoting the discretisation error as ${\boldsymbol \sigma} = (\sigma_x,\sigma_y,\sigma_z)$, the equation then takes the form of
\begin{align}
  \frac{1}{2Re} D_2 \tilde{\bf u}^{n+1} - \Big(\frac{1}{2Re} (\alpha^2+\gamma^2)  + \frac{2}{Re\Delta t} \Big) \tilde{\bf u}^{n+1}   - \nabla \tilde p^{n+1} & =-\tilde{\bf r}^n  + \boldsymbol\sigma    \label{helmholtzEq}
\end{align}
where $D_1,D_2$ are the first-order and second-order spatial derivatives for the Gauss-Lobatto points and the wave-like assumption in space has been adopted. We have overloaded the same notations $\tilde{\bf u}, \tilde p, \tilde{\bf r}$ to represent the unknowns or expressions in the physical and spectral space. Taking the divergence of the above equations and utilising the discretised continuity equation $i\alpha \tilde u^{n+1}+ D_1 \tilde v^{n+1}+ i\gamma \tilde w^{n+1}=0$, we arrive at
\begin{align}
-\nabla^2 \tilde p^{n+1} = -\nabla \cdot \tilde{\bf r}^{n} + \nabla \cdot \boldsymbol \sigma
\end{align}
Built on the idea of a Green's function, the influence matrix method decomposes the solution at the time step $n+1$ as follows
\begin{align}
\begin{pmatrix} \tilde p^{n+1} \\ \tilde {\bf u}^{n+1}\end{pmatrix}  = \begin{pmatrix} \tilde p_p \\ \tilde {\bf u}_p  \end{pmatrix} +  \alpha_{t}  \begin{pmatrix} {\tilde p}_t \\ \tilde {\bf u}_t \end{pmatrix}   + \alpha_{b}  \begin{pmatrix} \tilde p_b \\ \tilde {\bf u}_b  \end{pmatrix}   + \alpha^c_t   \begin{pmatrix} \tilde p^c_t \\ \tilde {\bf u}^c_t \end{pmatrix}   + \alpha^c_b   \begin{pmatrix} \tilde p^c_b \\  \tilde {\bf u}^c_b  \end{pmatrix}     \label{totaldecomp}
\end{align}
where the coefficients $\alpha_t, \alpha_b,\alpha_t^c, \alpha_b^c$ will be determined to satisfy the continuity condition and the boundary condition at the walls.

Among these components, $\tilde p_p , \tilde {\bf u}_p $ solve
\begin{subeqnarray}
-\nabla^2 \tilde p_p &=& -\nabla \cdot \tilde{\bf r}^{n}  \\
  \frac{1}{2Re} D_2 \tilde{\bf u}_p - \Big(\frac{1}{2Re} (\alpha^2+\gamma^2)  + \frac{2}{Re\Delta t} \Big) \tilde{\bf u}_p   - \nabla \tilde p_p & =&-\tilde{\bf r}^n
\end{subeqnarray}
with $\tilde p_p(\pm h)=0$ and $\tilde {\bf u}_p(\pm h)=\mathbf U_b$. 

The components $\tilde p_t, \tilde {\bf u}_t$ solve
\begin{subeqnarray}
-\nabla^2 \tilde p_t &=& 0  \\
  \frac{1}{2Re} D_2 \tilde{\bf u}_t - \Big(\frac{1}{2Re} (\alpha^2+\gamma^2)  + \frac{2}{Re\Delta t} \Big) \tilde{\bf u}_t   - \nabla \tilde p_t & =& \bf 0
\end{subeqnarray}
with $\tilde p_t(h)=1, \tilde p_t(-h)=0$ and $\tilde {\bf u}_t(\pm h)=\bf 0$. 

The components $\tilde p_b, \tilde {\bf u}_b$ solve
\begin{subeqnarray}
-\nabla^2 \tilde p_b &= &0  \\
  \frac{1}{2Re} D_2 \tilde{\bf u}_b - \Big(\frac{1}{2Re} (\alpha^2+\gamma^2)  + \frac{2}{Re\Delta t} \Big) \tilde{\bf u}_b   - \nabla \tilde p_b & =& \bf 0
\end{subeqnarray}
with $\tilde p_b(h)=0, \tilde p_b(-h)=1$ and $\tilde{\bf u}_b(\pm h)=\bf 0$. 

The components $\tilde p_t^c, \tilde{\bf u}_t^c$ solve
\begin{subeqnarray}
-\nabla^2 \tilde p_t^c &= &-D_1 \sigma_t \\
  \frac{1}{2Re} D_2 \tilde{\bf u}_t^c- \Big(\frac{1}{2Re} (\alpha^2+\gamma^2)  + \frac{2}{Re\Delta t} \Big) \tilde{\bf u}_t^c   - \nabla \tilde p^c_t & =&\bf 0
\end{subeqnarray}
with $\tilde p^c_t(\pm h)=0$ and $\tilde {\bf u}_t^c(\pm h)=\bf 0$. Note that $\sigma_y$ has been further decomposed as $\sigma_y=\alpha_t^c \sigma_t + \alpha_b^c \sigma_b $.

Finally, the components $\tilde p_b^c,\tilde{\bf u}_b^c$ solve
\begin{subeqnarray}
-\nabla^2 \tilde p_b^c &=& -D_1 \sigma_b  \\
  \frac{1}{2Re} D_2 \tilde{\bf u}_b^c- \Big(\frac{1}{2Re} (\alpha^2+\gamma^2)  + \frac{2}{Re\Delta t} \Big) \tilde{\bf u}_b^c   - \nabla \tilde p^c_b & =&\bf 0
\end{subeqnarray}
with $\tilde p^c_b(\pm h)=0$ and $\tilde{\bf u}_b^c(\pm h)=0$. 

After solving all the subsystems, one can assemble the solutions in Eq. (\ref{totaldecomp}) to determine the four coefficients $\alpha_t, \alpha_b,\alpha_t^c, \alpha_b^c$, requiring four equations. Two of the equations enforce the continuity equation at the boundaries, i.e., $D_1 \tilde v^{n+1}=0$ at $y=\pm h$ or by substitution
\begin{subeqnarray}
D_1 \tilde v_p + \alpha_t D_1 \tilde v_t + \alpha_b D_1 \tilde v_b + \alpha^c_t D_1 {\tilde v^c_t} + \alpha^c_b D_1 {\tilde v^c_b} &=&0,  \ \ \ \ \ \ \ \ \ \text{at} \ \  y=\pm h.
\end{subeqnarray}
The other two equations implement the boundary corrections, i.e., 
\begin{subeqnarray}
\frac{1}{2Re} D_2{\tilde v^{n+1}} - \Big(\frac{1}{2Re} (\alpha^2+\gamma^2)  + \frac{2}{Re\Delta t} \Big) \tilde{v}^{n+1}-   D_1 \tilde p^{n+1} &=& -\tilde r_y^n - \alpha^c_t,  \ \ \ \text{at} \ \  y=h, \\ 
\frac{1}{2Re} D_2{\tilde v^{n+1}} - \Big(\frac{1}{2Re} (\alpha^2+\gamma^2)  + \frac{2}{Re\Delta t} \Big) \tilde{v}^{n+1} -   D_1 \tilde p^{n+1}&= &-\tilde r_y^n - \alpha^c_b,  \ \ \ \text{at} \ \  y=-h.
\end{subeqnarray}
After these steps, we can obtain the pressure $\tilde p^{n+1}$ that enforces the continuity condition at all the grid points (including the boundary grid points), and the $\tilde v^{n+1}$ at the next time step. The other two velocity components, $\tilde u^{n+1}$ and $\tilde w^{n+1}$, can be obtained by solving the Helmholtz equations (\ref{helmholtzEq}). When solving these equations, the matrix diagonalisation method proposed by \citet{Ehrenstein1989} is employed, significantly enhancing the algorithm's efficiency. To ensure accurate computation of the nonlinear terms in physical space, the $3/2$ dealiasing rule is applied along the periodic directions.

\begin{figure}
	\centering
	\includegraphics[width=0.45\textwidth,trim= 100 0 100 50,clip]{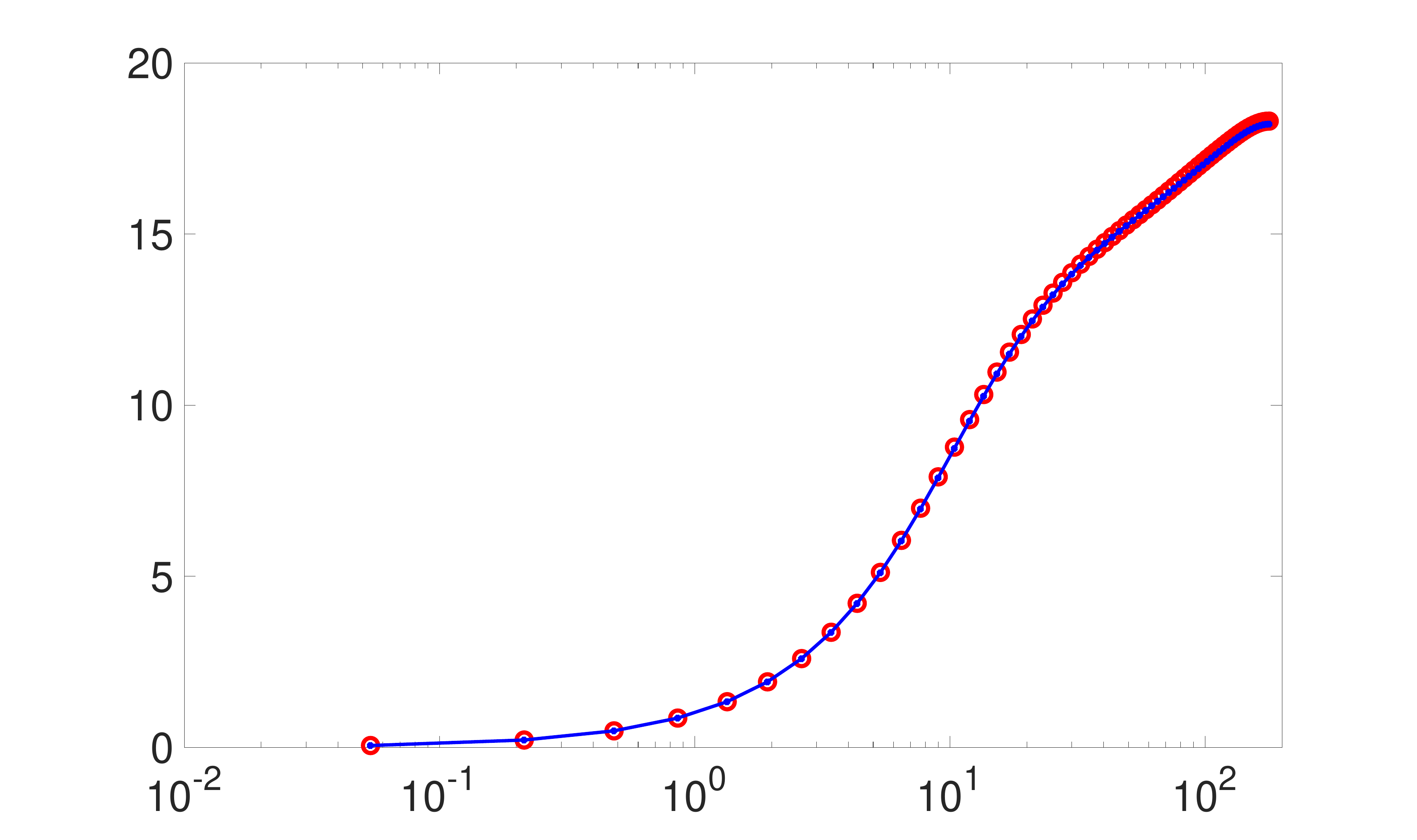}
	\put(-88,-5){\large$y^+$}
	\put(-180,55){\large$U^+$}			
	\includegraphics[width=0.45\textwidth,trim= 100 0 100 50,clip]{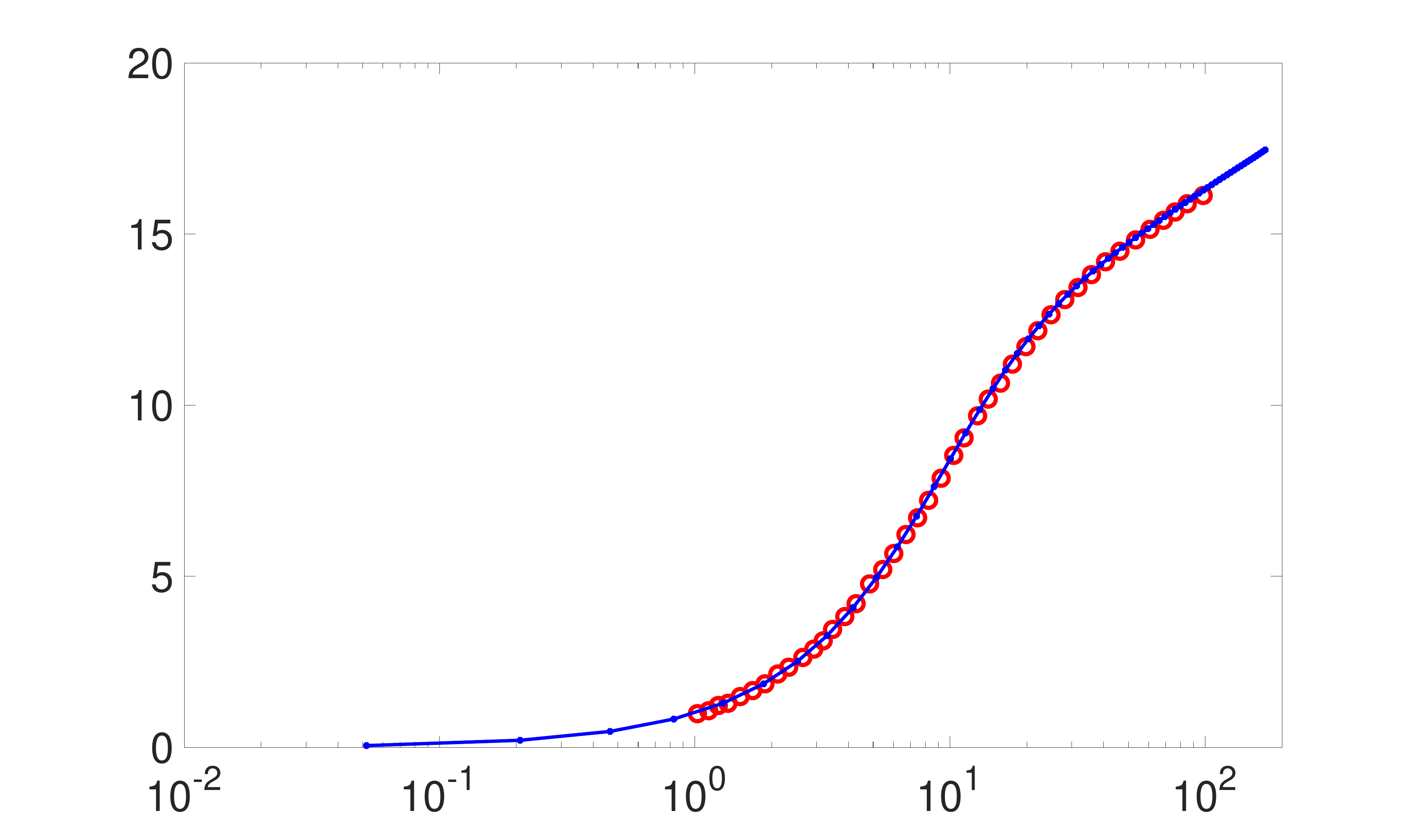}
	\put(-88,-5){\large$y^+$}	
	\\
	\includegraphics[width=0.45\textwidth,trim=100 0 100 50,clip]{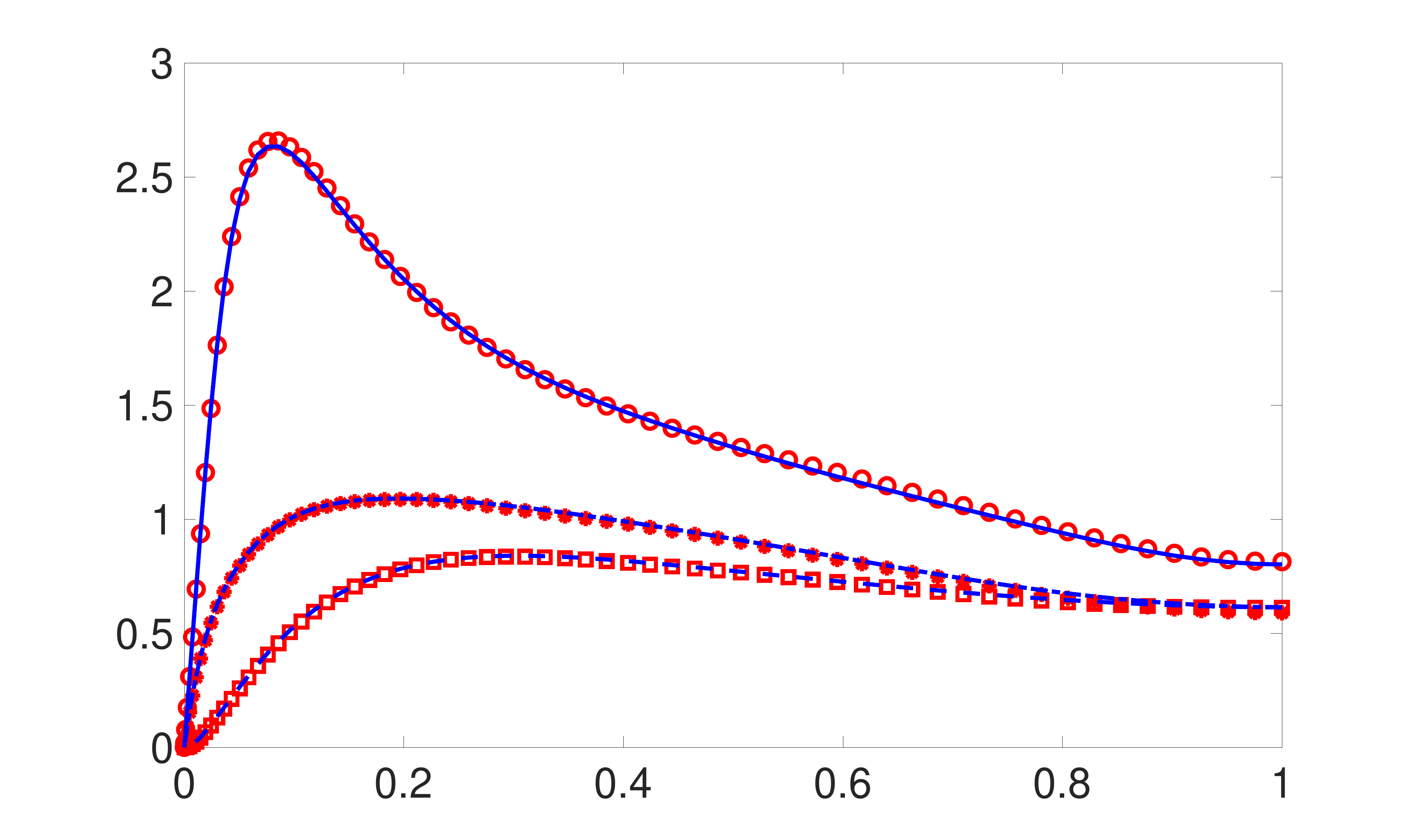}
	\put(-88,-5){\large$y$}
	\put(-190,75){\large$u^+_{rms}$}
	\put(-190,45){\large$w^+_{rms}$}
	\put(-190,25){\large$v^+_{rms}$}		
	\includegraphics[width=0.45\textwidth,trim=100 0 100 50,clip]{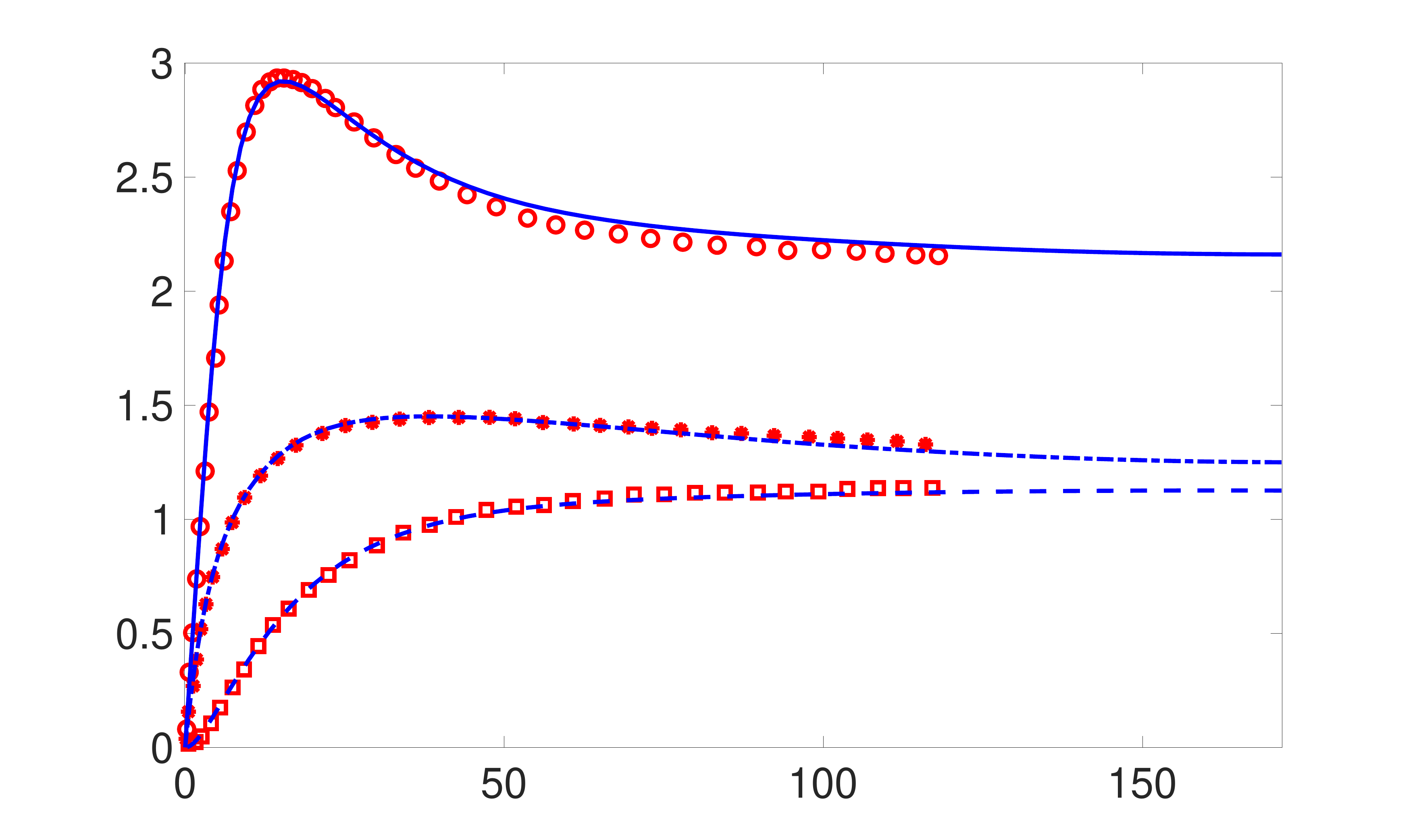}
	\put(-88,-5){\large$y^+$}
	\\
	\includegraphics[width=0.45\textwidth,trim=100 0 100 50,clip]{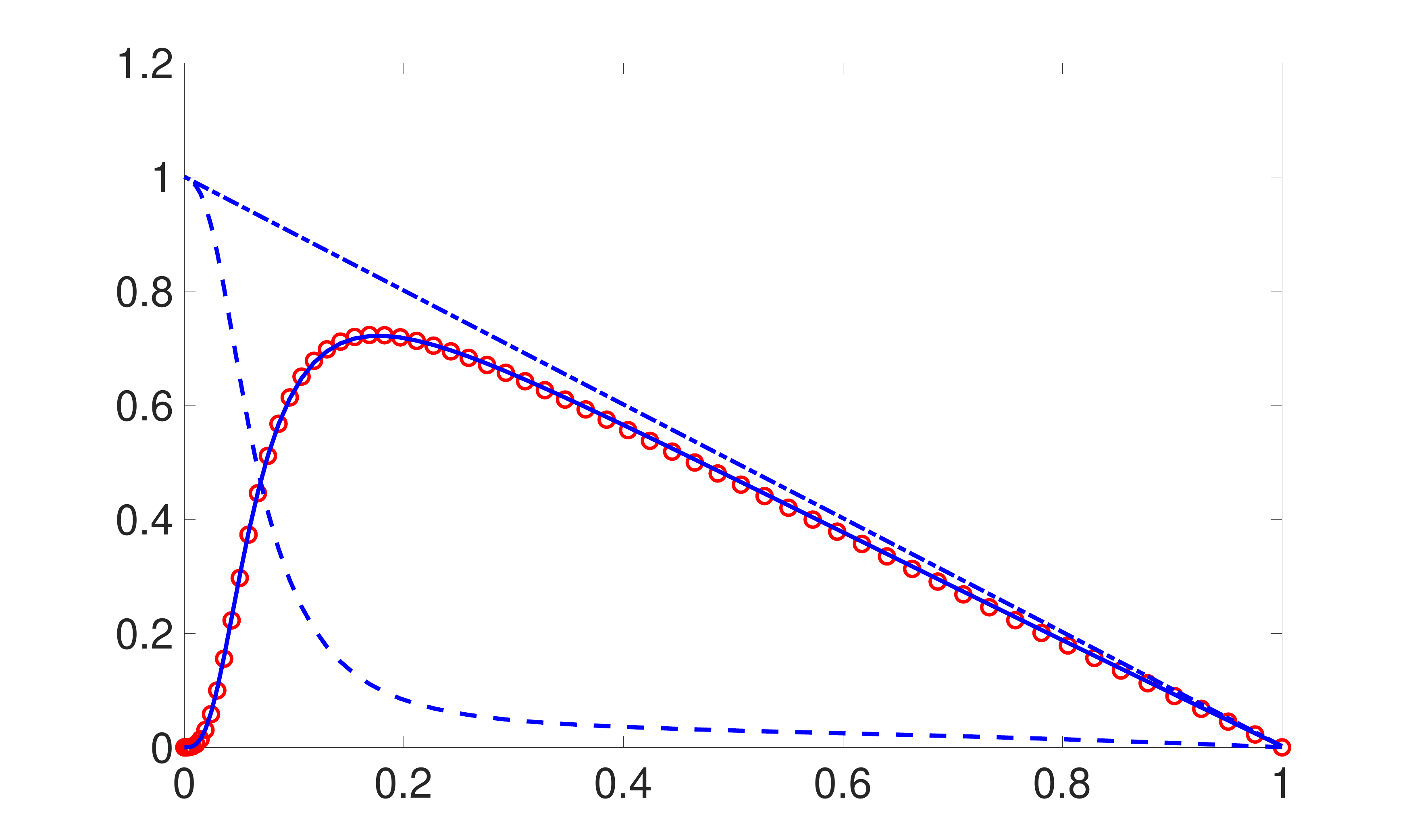}
	\put(-88,-5){\large$y$}
	\put(-185,55){\large${U^+}'$}		
	\put(-188,35){\large$RS^+$}	
	\includegraphics[width=0.45\textwidth,trim=100 0 100 50,clip]{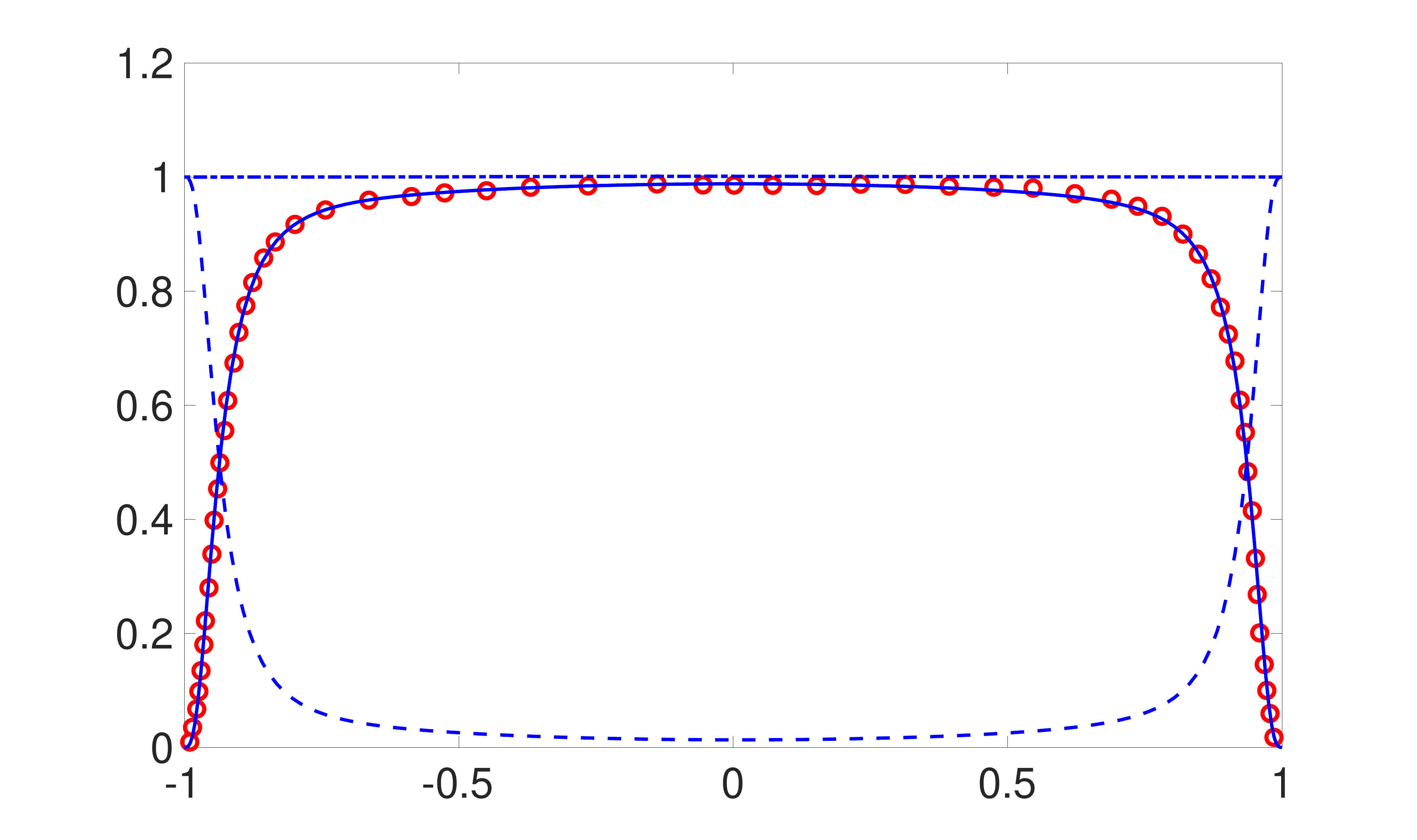}
	\put(-88,-5){\large$y$}
	\caption{Verification of the DNS code. Top row: mean velocity profiles; middle row: profiles of turbulence intensities; bottom row: profiles of shear stresses (solid lines: Reynolds stress ($RS^+$); dashed lines: viscous shear stress (${U^+}'$); dash-dotted lines: total stress). Left column: plane Poiseuille flow at $Re_\tau\approx 177.8$ compared to \cite{Moser1999} (red symbols); right column: plane Couette flow at $Re_\tau\approx 171.8$ compared to \cite{Lee1991} (red symbols), whose data points are manually extracted. Superscript $^+$ indicates normalisation with respect to the wall units. Here, $y,y^+$ denote the distance from a wall. }
	\label{veriDNS}
\end{figure}
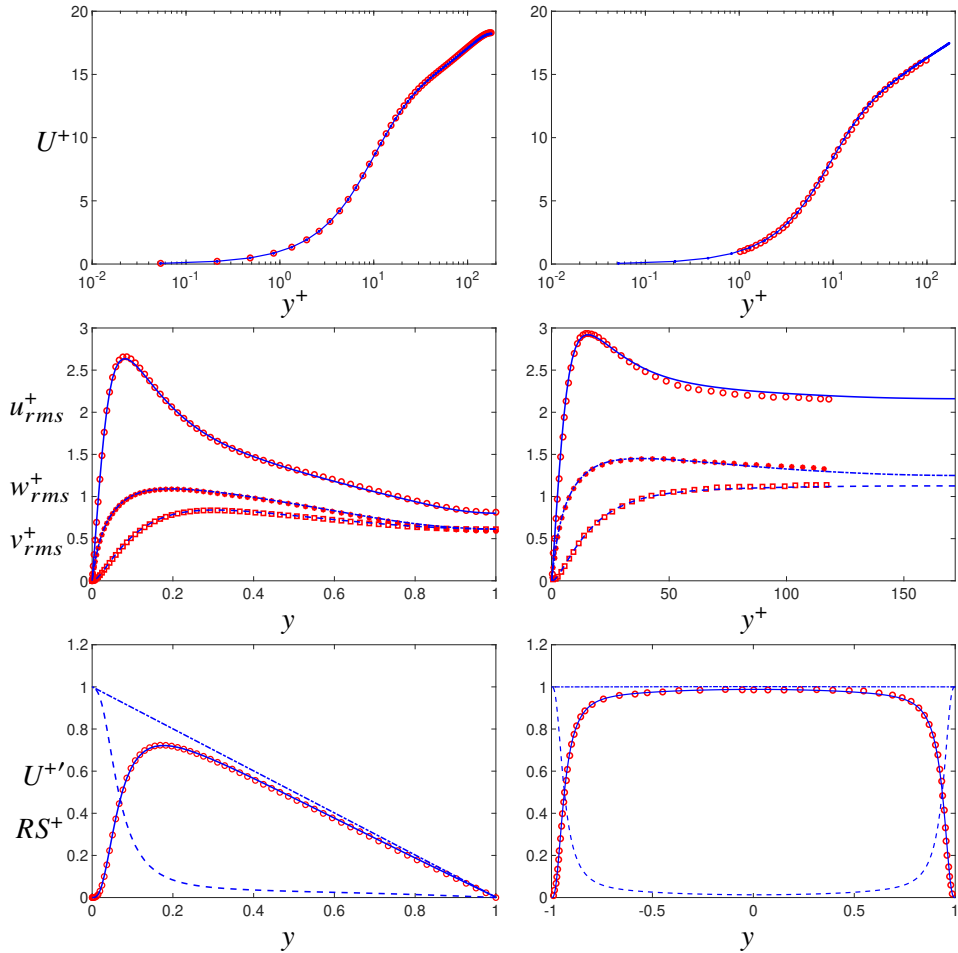

We present the validation of our numerical code in figure \ref{veriDNS} for turbulent channel flow and turbulent Couette flow. For the channel flow with constant mass flux at $Re_\tau \approx 180$, the computation domain is ($4\pi, 2,  \frac{4}{3}\pi$) with the resolution $129\times 129\times 89$ following the numerical setup in \cite{Moser1999}. The length is normalised by the channel half-height. For the plane Couette flow, it is well known that the computational domain significantly influences flow statistics due to its impact on resolving large-scale dynamics \citep{Komminaho1996,Pirozzoli2014}. We chose to validate our implementation against the pseudospectral simulations of \cite{Lee1991}, following their numerical domain ($4\pi, 2, \frac{8}{3}\pi$) and a similar resolution $129\times 129\times 193$ along the $x,y,z$ directions. As shown in the figure, our results exhibit a good agreement with the classic data.

\end{appendix}

\bibliographystyle{jfm}


\begin{thebibliography}{44}
\expandafter\ifx\csname natexlab\endcsname\relax\def\natexlab#1{#1}\fi
\def\au#1{#1} \def\ed#1{#1} \def\yr#1{#1}\def\at#1{#1}\def\jt#1{\textit{#1}}
  \def\bt#1{#1}\def\bvol#1{\textbf{#1}} \def\vol#1{#1} \def\pg#1{#1}
  \def\publ#1{#1}\def\arxiv#1{#1}\def\org#1{#1}\def\st#1{\textit{#1}}

\bibitem[Akhavan {\em et~al.\/}(1991{\natexlab{{\em a\/}}})Akhavan, Kamm \&
  Shapiro]{Akhavan1991}
{\sc \au{Akhavan, R.}, \au{Kamm, R.~D.} \& \au{Shapiro, A.~H.}}
  \yr{1991{\natexlab{{\em a\/}}}}  \at{{An investigation of transition to
  turbulence in bounded oscillatory Stokes flows Part 1. Experiments}}.  \jt{J.
  Fluid Mech.}  \bvol{225},  \pg{395--422}.

\bibitem[Akhavan {\em et~al.\/}(1991{\natexlab{{\em b\/}}})Akhavan, Kamm \&
  Shapiro]{Akhavan1991a}
{\sc \au{Akhavan, R.}, \au{Kamm, R.~D.} \& \au{Shapiro, A.~H.}}
  \yr{1991{\natexlab{{\em b\/}}}}  \at{{An investigation of transition to
  turbulence in bounded oscillatory Stokes flows Part 2. Numerical
  simulations}}.  \jt{J. Fluid Mech.}  \bvol{225},  \pg{423--444}.

\bibitem[Biau(2016)]{Biau2016}
{\sc \au{Biau, D.}} \yr{2016}  \at{{Transient growth of perturbations in Stokes
  oscillatory flows}}.  \jt{J. Fluid Mech.}  \bvol{794},  \pg{R4}.

\bibitem[Blennerhassett \& Bassom(2002)]{Blennerhassett2002}
{\sc \au{Blennerhassett, P.~J.} \& \au{Bassom, A.~P.}} \yr{2002}  \at{{The
  linear stability of flat Stokes layers}}.  \jt{J. Fluid Mech.}  \bvol{464},
  \pg{393--410}.

\bibitem[Blennerhassett \& Bassom(2006)]{Blennerhassett2006}
{\sc \au{Blennerhassett, P.~J.} \& \au{Bassom, A.~P.}} \yr{2006}  \at{The
  linear stability of high-frequency oscillatory flow in a channel}.  \jt{J.
  Fluid Mech.}  \bvol{556},  \pg{1--25}.

\bibitem[Blondeaux \& Vittori(1994)]{Blondeaux1994}
{\sc \au{Blondeaux, P.} \& \au{Vittori, G.}} \yr{1994}  \at{{Wall imperfections
  as a triggering mechanism for Stokes-layer transition}}.  \jt{J. Fluid Mech.}
   \bvol{264},  \pg{107--135}.

\bibitem[Blondeaux \& Vittori(2021)]{Blondeaux2021}
{\sc \au{Blondeaux, P.} \& \au{Vittori, G.}} \yr{2021}  \at{{Revisiting the
  momentary stability analysis of the Stokes boundary layer}}.  \jt{J. Fluid
  Mech.}  \bvol{919},  \pg{A36}.

\bibitem[Cowley(1987)]{Cowley1987a}
{\sc \au{Cowley, S.~J.}} \yr{1987}  \at{{High frequency Rayleigh instability of
  Stokes layers}}.  \bt{In {\em Stability of Time Dependent and Spatially
  Varying Flows\/} (ed. \ed{D.~L. Dwoyer \& M.~Y. Hussaini})},  \pg{p. 261}.
  \publ{Springer-Verlag, New York}.

\bibitem[Davis(1976)]{Davis1976}
{\sc \au{Davis, S.~H.}} \yr{1976}  \at{The stability of time-periodic flows}.
  \jt{Annu. Rev. Fluid Mech.}  \bvol{8},  \pg{57--74}.

\bibitem[Ebadi {\em et~al.\/}(2019)Ebadi, White, Pond \& Dubief]{Ebadi2019}
{\sc \au{Ebadi, A.}, \au{White, C.~M.}, \au{Pond, I.} \& \au{Dubief, Y.}}
  \yr{2019}  \at{Mean dynamics and transition to turbulence in oscillatory
  channel flow}.  \jt{J. Fluid Mech.}  \bvol{880},  \pg{864--889}.

\bibitem[Ehrenstein \& Peyret(1989)]{Ehrenstein1989}
{\sc \au{Ehrenstein, U.} \& \au{Peyret, R.}} \yr{1989}  \at{{A Chebyshev
  collocation method for the Navier--Stokes equations with application to
  double-diffusive convection}}.  \jt{Int. J. Numer. Methods Fluids}
  \bvol{9}~(4),  \pg{427--452}.

\bibitem[Farrell \& Ioannou(1996)]{Farrell1996}
{\sc \au{Farrell, B.~F.} \& \au{Ioannou, P.~J.}} \yr{1996}  \at{{Generalized
  Stability Theory. Part II: Nonautonomous Operators}}.  \jt{J. Atmos. Sci.}
  \bvol{53}~(14),  \pg{2041 -- 2053}.

\bibitem[Hall(1978)]{Hall1978}
{\sc \au{Hall, P.}} \yr{1978}  \at{{The linear stability of flat Stokes
  layers}}.  \jt{Proc. R. Soc. Lond.}  \bvol{359}~(1697),  \pg{151--166}.

\bibitem[Haller(2015)]{Haller2015}
{\sc \au{Haller, G.}} \yr{2015}  \at{Lagrangian coherent structures}.
  \jt{Annu. Rev. Fluid Mech.}  \bvol{47},  \pg{137--162}.

\bibitem[Hino {\em et~al.\/}(1983)Hino, Kashiwayanagi, Nakayama \&
  Hara]{Hino1983}
{\sc \au{Hino, M.}, \au{Kashiwayanagi, M.}, \au{Nakayama, A.} \& \au{Hara, T.}}
  \yr{1983}  \at{Experiments on the turbulence statistics and the structure of
  a reciprocating oscillatory flow}.  \jt{J. Fluid Mech.}  \bvol{131},
  \pg{363--400}.

\bibitem[Hino {\em et~al.\/}(1976)Hino, Sawamoto \& Takasu]{Hino1976}
{\sc \au{Hino, M.}, \au{Sawamoto, M.} \& \au{Takasu, S.}} \yr{1976}
  \at{Experiments on transition to turbulence in an oscillatory pipe flow}.
  \jt{J. Fluid Mech.}  \bvol{75}~(2),  \pg{193--207}.

\bibitem[Jensen {\em et~al.\/}(1989)Jensen, Sumer \& Freds{\o}e]{Jensen1989}
{\sc \au{Jensen, B.~L.}, \au{Sumer, B.~M.} \& \au{Freds{\o}e, J.}} \yr{1989}
  \at{{Turbulent oscillatory boundary layers at high Reynolds numbers}}.
  \jt{J. Fluid Mech.}  \bvol{206},  \pg{265--297}.

\bibitem[Jim{\'e}nez(2013)]{Jimenez2013}
{\sc \au{Jim{\'e}nez, J.}} \yr{2013}  \at{How linear is wall-bounded
  turbulence?}  \jt{Phys. Fluids}  \bvol{25}~(11),  \pg{110814}.

\bibitem[von Kerczek \& Davis(1974)]{Kerczek1974}
{\sc \au{von Kerczek, C.} \& \au{Davis, S.~H.}} \yr{1974}  \at{{Linear
  stability theory of oscillatory Stokes layers}}.  \jt{J. Fluid Mech.}
  \bvol{62}~(4),  \pg{753--773}.

\bibitem[Kern {\em et~al.\/}(2021)Kern, Beneitez, Hanifi \&
  Henningson]{Kern2021}
{\sc \au{Kern, J.~S.}, \au{Beneitez, M.}, \au{Hanifi, A.} \& \au{Henningson,
  D.~S.}} \yr{2021}  \at{{Transient linear stability of pulsating Poiseuille
  flow using optimally time-dependent modes}}.  \jt{J. Fluid Mech.}
  \bvol{927},  \pg{A6}.

\bibitem[Kim {\em et~al.\/}(1987)Kim, Moin \& Moser]{Kim1987}
{\sc \au{Kim, J.}, \au{Moin, P.} \& \au{Moser, R.}} \yr{1987}  \at{{Turbulence
  statistics in fully developed channel flow at low Reynolds number}}.  \jt{J.
  Fluid Mech.}  \bvol{177},  \pg{133--166}.

\bibitem[Komminaho {\em et~al.\/}(1996)Komminaho, Lundbladh \&
  Johansson]{Komminaho1996}
{\sc \au{Komminaho, J.}, \au{Lundbladh, A.} \& \au{Johansson, A.~V.}} \yr{1996}
   \at{{Very large structures in plane turbulent Couette flow}}.  \jt{J. Fluid
  Mech.}  \bvol{320},  \pg{259--285}.

\bibitem[Lee \& Kim(1991)]{Lee1991}
{\sc \au{Lee, M.~J.} \& \au{Kim, J.~.}} \yr{1991} {The structure of turbulence
  in a simulated plane Couette flow}.  \bt{In {\em In Eighth Symposium on
  Turbulent Shear Flows\/}},  \pg{pp. 5.3.1--5.3.6.} Technical University of
  Munich.

\bibitem[Lekien {\em et~al.\/}(2007)Lekien, Shadden \& Marsden]{Lekien2007}
{\sc \au{Lekien, F.}, \au{Shadden, S.~C.} \& \au{Marsden, J.~E.}} \yr{2007}
  \at{{Lagrangian coherent structures in n-dimensional systems}}.  \jt{J. Math.
  Phys.}  \bvol{48}~(6),  \pg{065404}.

\bibitem[Lozano-Dur{\'a}n {\em et~al.\/}(2021)Lozano-Dur{\'a}n, Constantinou,
  Nikolaidis \& Karp]{Lozano-Duran2021}
{\sc \au{Lozano-Dur{\'a}n, A.}, \au{Constantinou, N.~C.}, \au{Nikolaidis,
  M.-A.} \& \au{Karp, M.}} \yr{2021}  \at{Cause-and-effect of linear mechanisms
  sustaining wall turbulence}.  \jt{J. Fluid Mech.}  \bvol{914},  \pg{A8}.

\bibitem[Luchini \& Bottaro(2014)]{Luchini2014}
{\sc \au{Luchini, P.} \& \au{Bottaro, A.}} \yr{2014}  \at{Adjoint equations in
  stability analysis}.  \jt{Annu. Rev. Fluid Mech.}  \bvol{46}~(1),
  \pg{493--517}.

\bibitem[Luo \& Wu(2010)]{Luo2010}
{\sc \au{Luo, J.} \& \au{Wu, X.}} \yr{2010}  \at{{On the linear instability of
  a finite Stokes layer: Instantaneous versus Floquet modes}}.  \jt{Phys.
  Fluids}  \bvol{22}~(5),  \pg{054106}.

\bibitem[Madabhushi {\em et~al.\/}(1993)Madabhushi, Balachandar \&
  Vanka]{Madabhushi1993}
{\sc \au{Madabhushi, R.~K.}, \au{Balachandar, S.} \& \au{Vanka, S.P.}}
  \yr{1993}  \at{A divergence-free chebyshev collocation procedure for
  incompressible flows with two non-periodic directions}.  \jt{J. Comput.
  Phys.}  \bvol{105}~(2),  \pg{199--206}.

\bibitem[Manna {\em et~al.\/}(2015)Manna, Vacca \& Verzicco]{Manna2015}
{\sc \au{Manna, M.}, \au{Vacca, A.} \& \au{Verzicco, R.}} \yr{2015}
  \at{{Pulsating pipe flow with large-amplitude oscillations in the very high
  frequency regime. Part 2. Phase-averaged analysis}}.  \jt{J. Fluid Mech.}
  \bvol{766},  \pg{272--296}.

\bibitem[Merkli \& Thomann(1975)]{Merkli1975}
{\sc \au{Merkli, P.} \& \au{Thomann, H.}} \yr{1975}  \at{Transition to
  turbulence in oscillating pipe flow}.  \jt{J. Fluid Mech.}  \bvol{68}~(3),
  \pg{567--576}.

\bibitem[Mitran {\em et~al.\/}(2008)Mitran, Forest, Yao, Lindley \&
  Hill]{Mitran2008}
{\sc \au{Mitran, S.~M.}, \au{Forest, M.~G.}, \au{Yao, L.}, \au{Lindley, B.} \&
  \au{Hill, D.~B.}} \yr{2008}  \at{{Extensions of the Ferry shear wave model
  for active linear and nonlinear microrheology}}.  \jt{J. Non-Newtonian Fluid
  Mech.}  \bvol{154}~(2),  \pg{120--135}.

\bibitem[Monkewitz \& Bunster(1985)]{Monkewitz1985}
{\sc \au{Monkewitz, P.~A.} \& \au{Bunster, A.}} \yr{1985}  \at{{The Stability
  of the Stokes Layer: Visual Observations and Some Theoretical
  Considerations}}.  \bt{In {\em Stability o f Time Dependent and Spatially
  Varying Flows\/} (ed. \ed{D.~L. Dwoyer \& M.~Y. Hussaini})}.
  \publ{Springer-Verlag New York}.

\bibitem[Moser {\em et~al.\/}(1999)Moser, Kim \& Mansour]{Moser1999}
{\sc \au{Moser, R.~D.}, \au{Kim, J.} \& \au{Mansour, N.~N.}} \yr{1999}
  \at{Direct numerical simulation of turbulent channel flow up to
  {$Re_{\tau}$=590}}.  \jt{Phys. Fluids}  \bvol{11}~(4),  \pg{943--945}.

\bibitem[Pier \& Schmid(2021)]{Pier2021}
{\sc \au{Pier, B.} \& \au{Schmid, P.~J.}} \yr{2021}  \at{Optimal energy growth
  in pulsatile channel and pipe flows}.  \jt{J. Fluid Mech.}  \bvol{926},
  \pg{A11}.

\bibitem[Pirozzoli {\em et~al.\/}(2014)Pirozzoli, Bernardini \&
  Orlandi]{Pirozzoli2014}
{\sc \au{Pirozzoli, S.}, \au{Bernardini, M.} \& \au{Orlandi, P.}} \yr{2014}
  \at{{Turbulence statistics in Couette flow at high Reynolds number}}.  \jt{J.
  Fluid Mech.}  \bvol{758},  \pg{327--343}.

\bibitem[Schmid \& Henningson(2001)]{Schmid2001}
{\sc \au{Schmid, P.~J.} \& \au{Henningson, D.~S.}} \yr{2001} {\em {Stability
  and Transition in Shear Flows}\/}. {\em Applied Mathematical Sciences\/} .
  \publ{Springer Verlag, New York}.

\bibitem[Shadden(2012)]{Shadden2012}
{\sc \au{Shadden, S.~C.}} \yr{2012}  \at{Lagrangian coherent structures}.
  \bt{In {\em Transport and Mixing in Laminar Flows: From Microfluidics to
  Oceanic Currents\/} (ed. \ed{R.~Grigoriev})},  \pg{pp. 59--89.}
  \publ{Berlin: Wiley-VCH}.

\bibitem[Thomas {\em et~al.\/}(2010)Thomas, Bassom, Blennerhassett \&
  Davies]{Thomas2010}
{\sc \au{Thomas, C.}, \au{Bassom, A.~P.}, \au{Blennerhassett, P.~J.} \&
  \au{Davies, C.}} \yr{2010}  \at{{Direct numerical simulations of small
  disturbances in the classical Stokes layer}}.  \jt{J. Eng. Math.}
  \bvol{68}~(3),  \pg{327--338}.

\bibitem[Thomas {\em et~al.\/}(2011)Thomas, Bassom, Blennerhassett \&
  Davies]{Thomas2011}
{\sc \au{Thomas, C.}, \au{Bassom, A.~P.}, \au{Blennerhassett, P.~J.} \&
  \au{Davies, C.}} \yr{2011}  \at{{The linear stability of oscillatory
  Poiseuille flow in channels and pipes}}.  \jt{Proc. R. Soc. Lond.}
  \bvol{467}~(2133),  \pg{2643--2662}.

\bibitem[Thomas {\em et~al.\/}(2015)Thomas, Blennerhassett, Bassom \&
  Davies]{Thomas2015}
{\sc \au{Thomas, C.}, \au{Blennerhassett, P.~J.}, \au{Bassom, A.~P.} \&
  \au{Davies, C.}} \yr{2015}  \at{{The linear stability of a Stokes layer
  subjected to high-frequency perturbations}}.  \jt{J. Fluid Mech.}
  \bvol{764},  \pg{193--218}.

\bibitem[Thomas {\em et~al.\/}(2014)Thomas, Davies, Bassom \&
  Blennerhassett]{Thomas2014}
{\sc \au{Thomas, C.}, \au{Davies, C.}, \au{Bassom, A.~P.} \&
  \au{Blennerhassett, P.~J.}} \yr{2014}  \at{{Evolution of disturbance
  wavepackets in an oscillatory Stokes layer}}.  \jt{J. Fluid Mech.}
  \bvol{752},  \pg{543--571}.

\bibitem[Vittori \& Verzicco(1998)]{Vittori1998}
{\sc \au{Vittori, G.} \& \au{Verzicco, R.}} \yr{1998}  \at{Direct simulation of
  transition in an oscillatory boundary layer}.  \jt{J. Fluid Mech.}
  \bvol{371},  \pg{207--232}.

\bibitem[Weideman \& Reddy(2000)]{Weideman2000}
{\sc \au{Weideman, J.~A.} \& \au{Reddy, S.~C.}} \yr{2000}  \at{{A MATLAB
  Differentiation Matrix Suite}}.  \jt{ACM Trans. on Mathematical Software}
  \bvol{26}~(4),  \pg{465--519}.

\bibitem[Wu(1992)]{Wu1992}
{\sc \au{Wu, X.}} \yr{1992}  \at{{The nonlinear evolution of high-frequency
  resonant-triad waves in an oscillatory Stokes layer at high Reynolds
  number}}.  \jt{J. Fluid Mech.}  \bvol{245},  \pg{553--597}.

\end{thebibliography}

\end{document}